\documentclass[pdflatex,sn-nature]{sn-jnl}

\usepackage{times}%
\usepackage{graphicx}%
\usepackage{multirow}%
\usepackage{amsmath,amssymb,amsfonts}%
\usepackage{amsthm}%
\usepackage{mathrsfs}%
\usepackage[title]{appendix}%
\usepackage{xcolor}%
\usepackage{textcomp}%
\usepackage{manyfoot}%
\usepackage{booktabs}%
\usepackage{algorithm}%
\usepackage{algorithmicx}%
\usepackage{algpseudocode}%
\usepackage{listings}%
\usepackage{hyperref}%
\newcommand\mnras{Mon. Not. R. Astro. Soc.}             

\theoremstyle{thmstyleone}%
\newcommand{\eg}{e.g.,~}
\newcommand{\ie}{i.e.,~}

\theoremstyle{thmstyletwo}%

\theoremstyle{thmstylethree}%

\begin{document}

\title[Future ability to test theories of gravity]{The future ability to
  test theories of gravity with black-hole shadows}


\author*[1]{\fnm{Akhil} \sur{Uniyal}}\email{akhil\_uniyal@sjtu.edu.cn}

\author[1]{\fnm{Indu} \sur{K. Dihingia}}

\author[1,2,3]{\fnm{Yosuke} \sur{Mizuno}}\email{mizuno@sjtu.edu.cn}

\author[3,4,5]{\fnm{Luciano} \sur{Rezzolla}}\email{rezzolla@itp.uni-frankfurt.de}

\affil[1]{\orgdiv{Tsung-Dao Lee Institute}, \orgname{Shanghai Jiao Tong
    University}, \orgaddress{\street{Shengrong Road 520},
    \postcode{201210}, \state{Shanghai}, \country{People's Republic of
      China\\}}}

\affil[2]{\orgdiv{School of Physics and Astronomy}, \orgname{Shanghai
    Jiao Tong University}, \orgaddress{\street{800 Dongchuan Road},
    \postcode{200240}, \state{Shanghai}, \country{People's Republic of
      China\\}}}

\affil[3]{\orgdiv{Institut f\"{u}r Theoretische Physik},
  \orgname{Goethe-Universit\"{a}t Frankfurt},
  \orgaddress{\street{Max-von-Laue-Strasse 1}, \postcode{D-60438},
    \state{Frankfurt am Main}, \country{Germany\\}}}

 \affil[4]{\orgdiv{CERN, Theoretical Physics Department},
  \orgaddress{1211 Geneva 23, Switzerland\\}}

\affil[5]{\orgdiv{School of Mathematics}, \orgname{Trinity College},
  \orgaddress{Dublin 2, Ireland}}


%
\abstract{The horizon-scale images of supermassive black holes (BHs) by
  the Event Horizon Telescope Collaboration (EHT) have provided new
  opportunities to test general relativity and other theories of
  gravity. In view of future projects, such as the next-generation Event
  Horizon Telescope (ngEHT) and the Black-Hole Explorer (BHEX), having
  the potential of enhancing our ability to probe extreme gravity, it is
  natural to ask: \textit{how much can two black-hole images differ?} To
  address this question and assess the ability of these projects to test
  theories of gravity with black-hole shadows, we use
  general-relativistic magnetohydrodynamic and radiative-transfer
  simulations to investigate the images of a wide class of accreting BHs
  deviating from the Kerr solution. By measuring the mismatch between
  images of different BHs we show that future missions will be able to
  distinguish a large class of BHs solutions from the Kerr solution when
  the mismatch in the images exceeds values between $2\%$ and $5\%$
  depending on the image-comparison metric considered. These
    results indicate future horizon-scale imaging with percent-level
    image fidelity can place meaningful observational constraints on
    deviations from the Kerr metric and thereby test strong-field
    predictions of general relativity.}



\maketitle

Black holes (BHs) are a fundamental prediction of general relativity (GR)
and are characterized by the peculiar existence of an event horizon, a
null 2-surface from which not even light can escape. A number of
astronomical observations have provided evidence of the existence of BHs,
either through the detection of gravitational
waves~\cite{LIGOScientific:2016aoc}, or through the dynamics of stars
around the center of our galaxy~\cite{GRAVITY:2018ofz}. The Event Horizon
Telescope (EHT) Collaboration has recently published a series of papers
providing the first-ever images of accreting supermassive BHs either in
the center of the M87 galaxy~\cite{EventHorizonTelescope:2019dse,
  EventHorizonTelescope:2019ths, EventHorizonTelescope:2019ggy, 
  EventHorizonTelescope:2021bee, EventHorizonTelescope:2024dhe}, or at
the center of the Milky Way~\cite{EventHorizonTelescope:2022wkp, 
EventHorizonTelescope:2022apq, EventHorizonTelescope:2022wok, 
EventHorizonTelescope:2022exc, EventHorizonTelescope:2024hpu}.

While all of this information provides convincing evidence that BHs as
predicted by GR represent the simplest and most natural interpretation of
all the collected data, the uncertainties in the measurement still leave
room for a number of alternative interpretations (see, \eg the discussion
in~\cite{Kocherlakota2021b}).
We recall that, in the presence of an emitting region, GR predicts that
the image of such region will consist of a series of nested ring-like
images where each ring is distinguished by the number of half-orbits that
photons make before reaching the observer. The limiting ring in this
sequence, and thus innermost ring, is also referred to as the ``photon
ring'' or ``$n \to \infty$ image''~\cite{Bardeen:1973tla,
  Luminet:1979nyg, Gralla:2019xty, Johnson:2019ljv,
  Kocherlakota:2024hyq}. We should note that this nomenclature
  is admittedly confusing. First, the rings in the sequence of
  higher-order images are themselves sometime referred to as ``photon
  rings'' although the ``photon ring'' is only the innermost of the
  photon rings (see, \eg \cite{Kocherlakota:2024hyq} for a
  discussion). Second, the photon ring should not be confused with the
  ``light ring'', which marks the location of the unstable circular orbit
  in a spherically symmetric spacetime in the absence of an emitting
  region (see, \eg \cite{difilippo2025} for a discussion).  Hence, the
accurate measurement structure of the photon rings and of the location of
of the location and represents the most compelling route to investigate
gravity in the regime of strong but stationary curvature.  Given these
considerations, it is natural to ask if two BH images will be different
and if so how much they will actually differ.

In 2018, before the EHT Collaboration had revealed the first image of
M87$^*$, we considered this question and explored what was then the
``current'' ability to test theories of gravity with BH
shadows~\cite{Mizuno:2018lxz}. The conclusion drawn at that time when
comparing a Kerr BH with a dilaton was that, with the nominal angular
resolution of $\sim 20\, \mu{\rm as}$ for the EHT telescopes,
distinguishing the two BHs was extremely challenging. Seven years later,
a number of very large baseline interferometry (VLBI) projects with a
significant increase in angular resolution are
planned~\cite{Akiyama:2024msp, Kawashima:2024svy}, either with
Earth-based projects, such as the next-generation Event Horizon Telescope
(ngEHT)~\cite{Johnson2023}, or with space-based VLBI
missions~\cite{Roelofs2019, Fromm2021, Gurvits2021, Gurvits2022}, such
the Black Hole Explorer (BHEX)~\cite{Ayzenberg:2023hfw, Akiyama:2024msp,
  Kawashima:2024svy}. The goal of all these projects is to bring the
angular resolution to a few $\mu{\rm as}$~\cite{Johnson:2024ttr,
  Lupsasca:2024xhq, Marrone2024}. The goal of our work is to assess the
ability of these projects (and of future ones) to distinguish BH images.

Given the complex dynamics of plasma and radiation in the vicinity of an
accreting BH, general-relativistic magnetohydrodynamical (GRMHD)
simulations and the use of general-relativistic radiation-transfer (GRRT)
rendering represent the best route address this question. Initial works
have already provided important input on this issue and shown that the
prospects of distinguishing ultra-compact objects, \eg boson stars,
wormholes, gravastars, and naked singularities, appear
optimistic~\cite{Olivares2020, EventHorizonTelescope:2022xqj,
  Combi:2024ehi, Mishra:2024bpl, Dihingia:2024cch, Cemeljic:2025bqz},
black-hole spacetimes are far harder to distinguish~\cite{Mizuno:2018lxz}
and a general degeneracy problem needs to be
addressed~\cite{Kocherlakota:2022jnz}. While some work in this direction
has already been started~\cite{Mizuno:2018lxz,
  EventHorizonTelescope:2022xqj, Nampalliwar:2022smp, Chatterjee:2023rcz,
  Chatterjee:2023wti, Roder:2023oqa, Jiang:2024vgn}, it also clear that
performing simulations for all different gravity theories and comparing
them with the observation is not computationally feasible.

Fortunately, it is possible to explore a very large of the possible space
of parameters by using the parametric Konoplya-Rezzolla-Zhidenko (KRZ)
metric~\cite{Konoplya:2016jvv, Younsi2016} that has been shown to provide
an accurate representation of broad class of axisymmetric and stationary
black-hole spacetimes. More specifically, we adopt the subclass of KRZ
spacetimes that leads to a separable form of the Hamilton-Jacobi
equations and that is written in horizon-penetrating
coordinates~\cite{Konoplya2018, Ma:2024kbu}. In this way, we report the
results of GRMHD and GRRT simulations of three significantly different
KRZ BH spacetimes having identical initial conditions and compare the
corresponding images in terms of mismatches from the Kerr solution.

\section{Results}
\label{sec2}

To obtain realistic black-hole images, we first perform three-dimensional
(3D) GRMHD simulations employing the \texttt{BHAC}
code~\cite{Porth:2016rfi, Olivares:2019dsc} of magnetized accretion flows
onto a BH, which is either a Kerr BH in GR or a ``KRZ BH'', \ie a BH
resulting from a given choice of the KRZ parameters, where the latter are
chosen so as to select specific ``corner-cases'' in the space of
parameters. We note that while all BHs considered have the same
dimensionless spin $J/M^2:=a=0.9375$ (here $J$ and $M$ are the BH angular
momentum and mass and a high spin is chosen to enhance the strong-field
effects, see Supplementary Information Section A and
  Supplementary Fig. 1), they inevitably differ in one (or
more) physical properties, \eg the position of the innermost stable
circular orbit (ISCO), of the unstable circular photon orbits, or of the
event horizon (Ref.~\cite{Mizuno:2018lxz} has shown that fixing
  the ISCO or the unstable circular photon orbits yields very similar
  results). Given these differences, we set the Kerr and KRZ BHs to have
also all the same position of the event horizon, which we place at
$r_h=1.348\,r_{g}$, where $r_{g}=M$ is the gravitational radius ($G=c=1$
here). In addition, because the initial tori around the
BH~\cite{Fishbone1976RelativisticFD, Uniyal:2024sdv} will also have
different sizes for the same choice of specific angular momentum, we tune
the latter so that and the position of the inner edge is fixed and they
all have the same total rest-mass (see~\nameref{methods} for details on
the simulations).

\begin{figure*}
    \centering \includegraphics[width=0.98\textwidth]{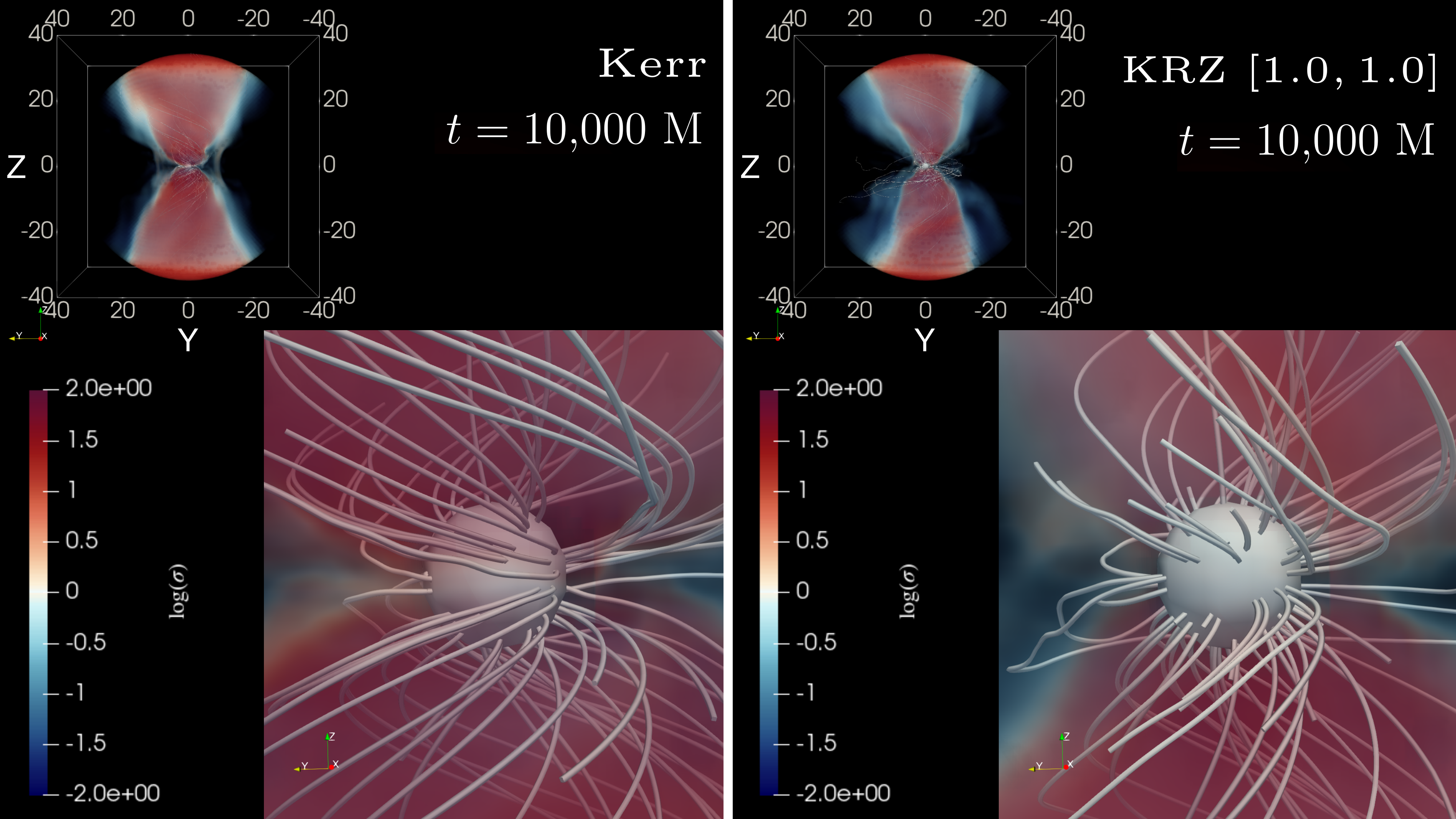}
    \caption{\textbf{Volume rendering image for Kerr and KRZ [1.0,1.0] BHs.} Visualisation of the magnetic-field lines and volume
      rendering of the magnetisation $\sigma$ at $t=10,\!000 \, M$ for a
      GRMHD simulation of an accretion onto a Kerr BH with $a_0=1=a_1$
      (left panel) or a KRZ BH (right). Shown on the top left of the two
      panels is reported a large-scale view.}
    \label{fig:3D-MAD}
\end{figure*}

As a representative example, we show in Fig.~\ref{fig:3D-MAD} the
volume-rendering of the magnetisation $\sigma := B^2/\rho$, which
compares the magnetic energy density($\propto B^2$, with $B$ the magnetic
field) the rest-mass energy density (with $\rho$ the rest-mass density),
together with representative magnetic-field lines (the insets show a
large-scale view). The snapshots are taken at time $t=10,\!000\,M$ and
are representative of a Kerr BH (left) or a KRZ BH that has the largest
deviations from the Kerr solution (\ie $a_0=1=a_1$). We note that we
could have chosen even larger values of $a_0$ and $a_1$, but have
restricted our attention to $a_0\leq 1,\,a_1\leq 1$ because these
coefficients are expected to be small and because larger values would
yield spacetimes that are significantly different and hence easier to
distinguish.

\begin{figure}
  \centering
  \includegraphics[width=0.5\textwidth]{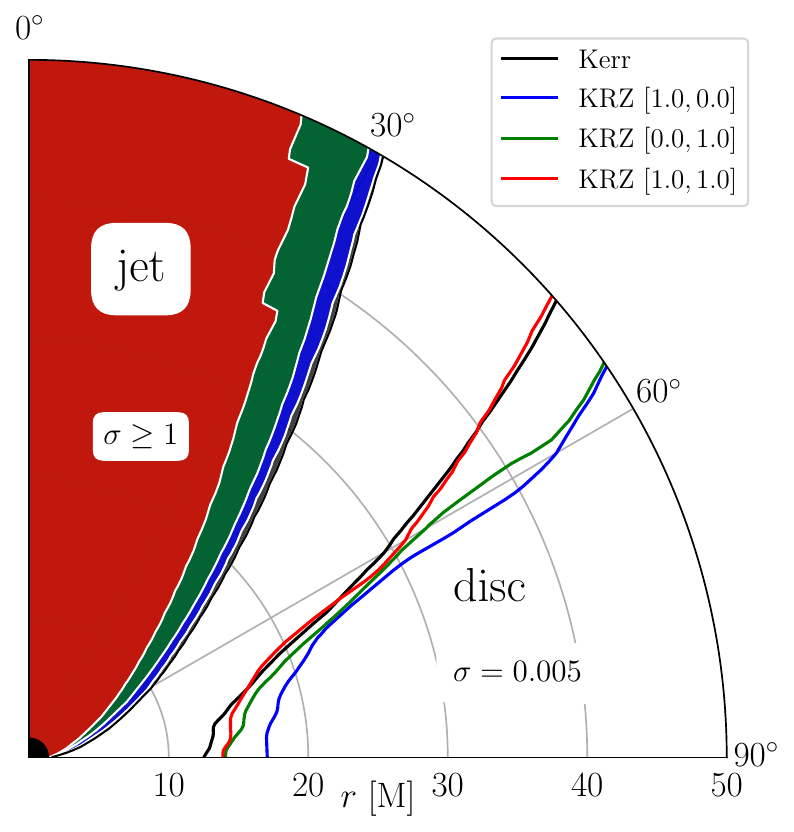}
  \caption{\textbf{$\sigma$ Contours for Kerr and KRZ BHs.} $\sigma=1.0$
    contours for simulations onto a Kerr BH (black) and three
    representative KRZ BHs (blue, green, and red). The values refer to
    data that is azimuthally- and time-averaged ($t=8,\!000-10,\!000 \,
    M$). The jet region is described by $\sigma \geq 1$, while
      the disc region has $\sigma \ll 1$. }
\label{fig:polar_3D}
\end{figure}

As typical in GRMHD simulations of accretion onto BHs, accretion is
triggered by the development the magnetorotational instability (MRI) that
leads to a turbulent plasma dynamics and outward transport of angular
momentum. Because of accretion, magnetic field accumulates on the
black-hole horizon, creating a highly magnetised region around the
rotation axis of the BH and a corresponding outflow that is referred to
as the ``jet''. The qualitative contrast offered in Fig.~\ref{fig:3D-MAD}
shows that the accretion flow in the case of the KRZ BH is more turbulent
and that the magnetic-field lines are more tightly packed near the
horizon. A more quantitative comparison is reported in
Fig.~\ref{fig:polar_3D}, which shows the time- and azimuthally-averaged
($t=8,\!000-10,\!000\,M$) polar contours of the $\sigma=1.0$, which are
often used to mark the edge of the jet, either for a Kerr BH (black line)
or for three different KRZ BHs representing corner cases in the space of
parameter defined by the coefficients $a_0, a_1$ (first and second number
in the square brackets). Also reported are the contour lines
corresponding to $\sigma=0.005$ and hence nominally representative of the
location of the disc. Note that the difference here are somewhat larger,
but the various lines are still very close to each other.

Tracking the contours of the magnetisation, it is then clear that while
the KRZ BHs tend to have systematically narrower jets, the actual
geometrical differences are rather small. Similar considerations apply
also for the time-averaged jet power calculated at $r = 50\,M$
(see~\nameref{methods}), which is $P_{\rm jet} = 7.30$ (in code units)
for a Kerr BH and $P_{\rm jet} = 13.21, 6.74$, and $5.24$ for the KRZ
BHs, where we note that the variation in the jet power is not as
systematic as in the jet section, where KRZ BHs are always smaller than
the corresponding Kerr BH. Hence, also from an energetic point of view,
despite the differences in the spacetimes, the actual energy losses are
overall very similar (see Supplementary Information Section B
  and Supplementary Figs. 2--4). These results anticipate the magnitude
of the differences that are to be expected from the horizon-scale images
-- including images of the jet base -- that will be made by future
EHT/ngEHT and BHEX experiments.

\begin{figure*}
  \centering
    \includegraphics[width=0.98\textwidth]{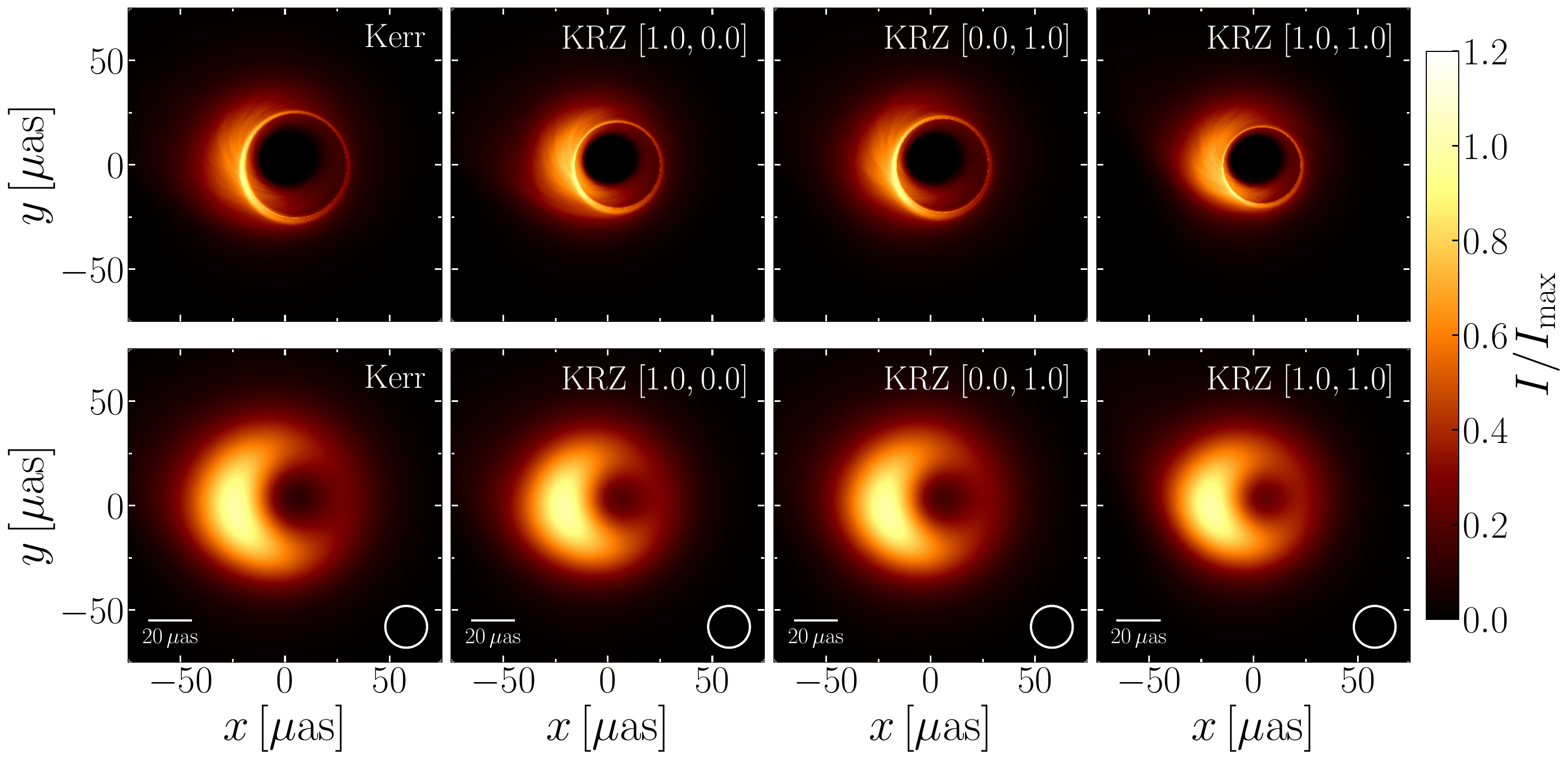}
    \caption{\textbf{$230\,{\rm GHz}$ time-averaged images: Kerr vs. KRZ
        BHs.} Time-averaged ($t=8,\!000-10,\!000\,M$) intensity images at
      $230\,{\rm GHz}$ of accretion flows onto a Kerr BH (left column)
      and onto three representative Kerr BHs (second to fourth
      column). While the top row reports the data from the simulation,
      the bottom one show images using a circular Gaussian beam with a
      full width at half maximum (FWHM) of $20\,\mu{\rm as}$. The
      inclination angle is $i=30^\circ$, the BH spin is pointing upwards,
      and the disc rotating counter-clock wise.}
    \label{fig:grrt}
\end{figure*}

Of course, the way in which the observations will provide information on
the BHs is via horizon-scale imaging and to this scope we have generated
GRRT images at a frequency of $230\, {\rm GHz}$ for a large number of BHs
with properties similar Sgr~A$^*$, namely, with a mass $M=4.14 \times
10^6\,M_{\odot}$, a distance of $8.127\,{\rm kpc}$ and where the
mass-accretion rate is set to be comparable to the mean total flux
density of $\sim 2.5\,{\rm Jy}$.

It is important to note that the choice of considering Sgr~A$^*$ as
reference source over the alternative supermassive BH M87$^*$ has
advantages and disadvantages. More specifically, in the case of Sgr~A$^*$
we know the mass to a precision of about $\sim 1\%$ but its emission is
subject to an intrinsic variability over a timescale of tens of minutes
and is affected by scattering as it propagates from the galactic center
towards the telescopes. By contrast, the mass of M87$^*$ is known to a
much smaller precision of only $\sim 20\%$, but the variability is on
timescales of days and the impact of scattering minimal. Because the
accurate measurement of the location of the light ring plays a crucial
role in any testing of theories of gravity, the advantages of a precise
knowledge of the mass dominate of the disadvantages induced by
variability and scattering; mass measurements of M87$^*$ with
  comparable precision of $\sim 1\%$ are expected with future
  experiments~\cite{Gurvits2022, Johnson2023, Kawashima:2024svy}. In
addition, the latter are expected to be considerably moderated by long
and repeated exposures, which will allow for the stable and persistent
position of the light ring to emerge over the variable emission.

The results of our analysis are summarised in Fig.~\ref{fig:grrt}, which
reports time-averaged shadow images of a Kerr BH (left column) and of the
three representative KRZ BHs. For each BH, the top row refers to the
GRMHD simulations, while the bottom one to observations using a circular
Gaussian beam with a full width at half maximum (FWHM) of $20\,\mu{\rm
  as}$ (the images refer to an inclination of $i=30^\circ$ but similar
considerations apply to any inclination).

\begin{figure}
  \centering
  \includegraphics[width=0.9\textwidth]{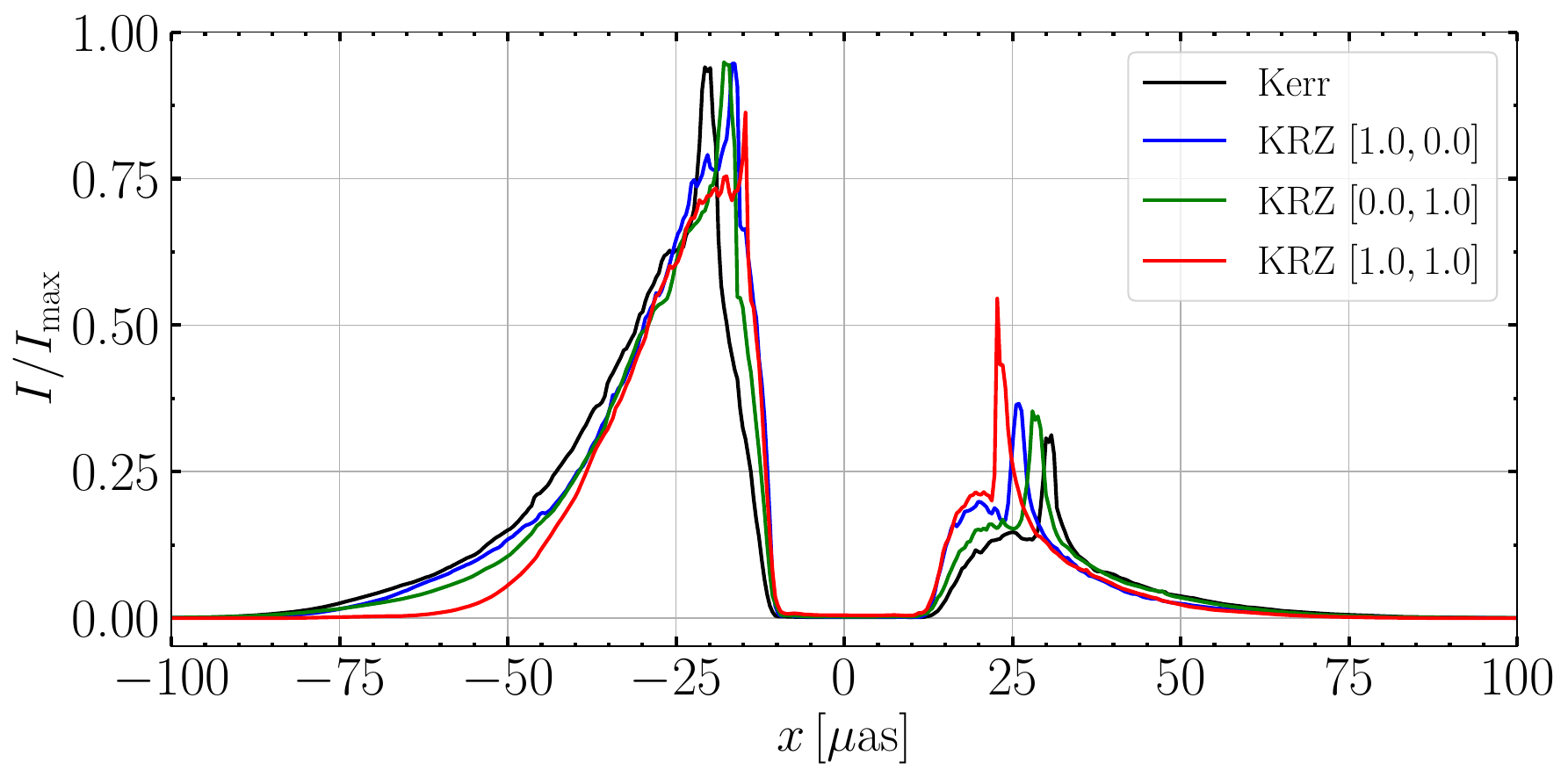}
  \caption{\textbf{Intensity profiles along $y=0$ for Kerr and KRZ BHs.} One-dimensional cuts at $y=0$ of the intensities reported in
    Fig.~\ref{fig:grrt} for the four BHs considered.}
  \label{fig:cross-cut}
\end{figure}

Clearly, all BH images show a similar morphology: a bright central photon
ring with extended ring-like emission coming from accretion flows near
the BH. The left side is brighter than the right side due to the Doppler
boosting of the plasma orbiting around the BH (we assume the disc to
rotate in the counter-clock direction and the spin to be pointing
upwards) Comparing the different images it is possible to appreciate a
difference in the photon-ring size and, in particular, the KRZ BH with a
``pure-$a_0$'' deformation has a markedly smaller shadow (\ie intensity
depression) than that of the Kerr BH (the relative differences
  in shadow size, photon-ring location, and ring total intensity with
  respect to the Kerr BH are $\sim 17\%, 15\%$ and $6\%$,
  respectively). On the other hand, the KRZ BH with a ``pure-$a_1$''
deformation exhibits a comparable shadow size (the differences
  in this case are $\sim 12\%, 9\%$ and $8\%$), while a KRZ BH with a
``mixed-$a_0$-$a_1$'' deformation shows a ring size that is both the
brightest and the smallest (the differences are $\sim 27\%,
  26\%$ and $9\%$). Overall, therefore, differences are obviously
present in the images but these are also rather small even when
considering the most extreme deviations from a Kerr BH. For a more
quantitative comparison, we show in Fig.~\ref{fig:cross-cut} a cross-cut at
$y=0$ of the intensity distributions reported in Fig.~\ref{fig:grrt}. This
representation highlights more clearly what already discussed above,
namely, that the KRZ BHs tend to have systematically smaller shadows than
for a Kerr BH and that the BH with a mixed-$a_0$-$a_1$ deformation has
the smallest shadow and that a BH with ``pure-$a_1$'' deformation
exhibits a shadow size that is comparable with that of a Kerr BH.

While it is reassuring to see that sizeable differences appear in the
shadow size, the question to address is whether such differences can be
measured by present and future experiments with their inevitable limits
on resolution. To this scope we consider a number of image-comparison
metrics that quantify pixel-by-pixel the difference between two images
(see~\nameref{methods} for details). Out of the various ones that we have
considered, we here report the results in terms of the normalised
cross-correlation coefficient (nCCC), which can be assimilated to an
``overlap'' between two images, so that ${\rm nCCC}=1$ refers to two
identical images and ${\rm nCCC}=0$ to two distinct images.

\begin{figure}
  \centering
  \includegraphics[width=0.45\textwidth]{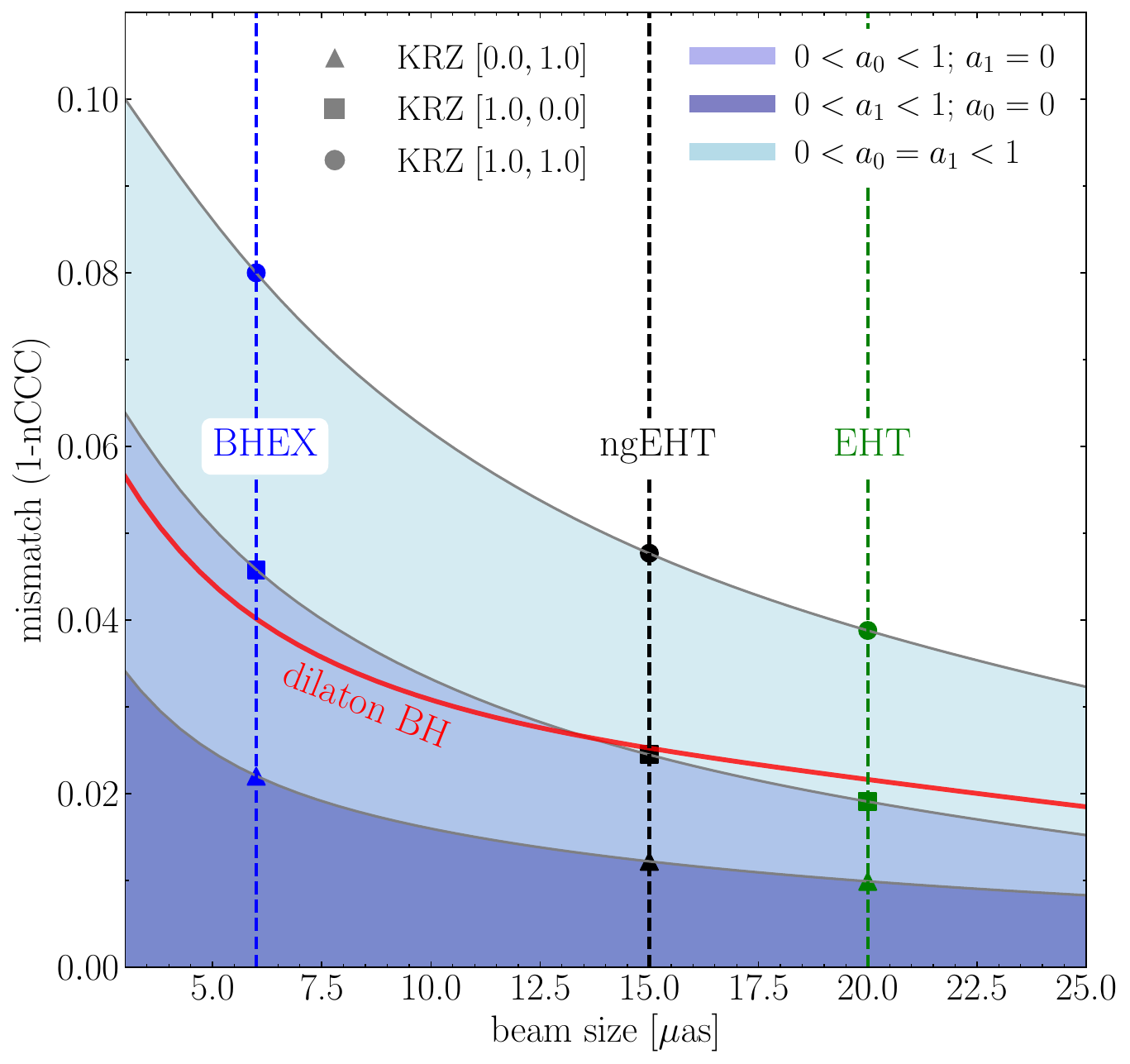}
  \hspace{0.5cm}
  \includegraphics[width=0.45\textwidth]{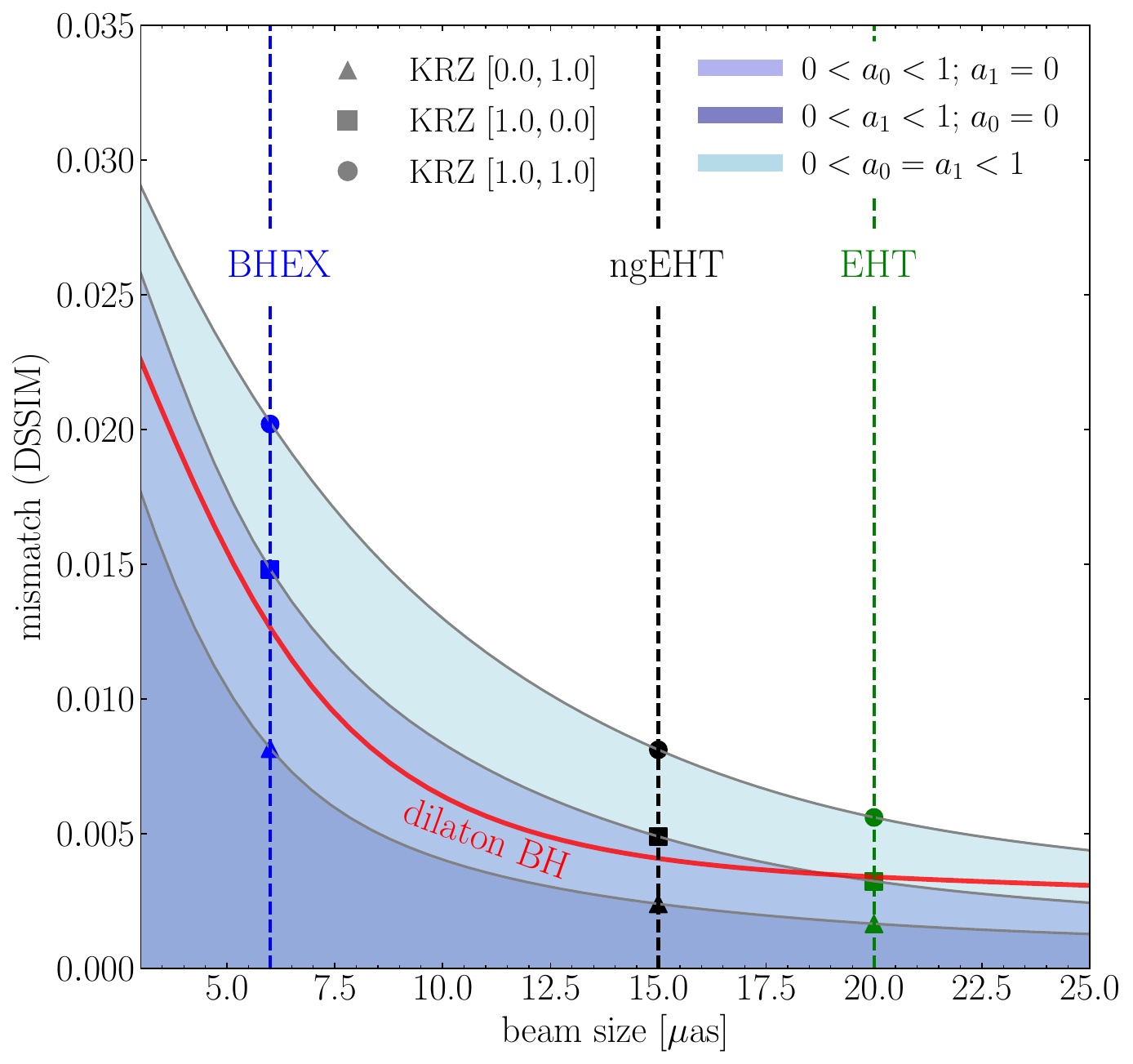}
  \caption{\textbf{Image-Comparison metrics as a function of beam size.} \textit{Left panel:} Image-comparison metric in terms of the
    ``mismatch'' $1-{\rm nCCC}$ for different beam sizes and KRZ BHs. The
    various colour-shaded regions show the variations of the mismatch
    between a Kerr BH and KRZ BH with properties set by the given
    colormap.  Reported with vertical lines are the present EHT
    resolutions and the expected ones for ngEHT and BHEX, while shown
    with different symbols are the mismatches corresponding to the three
    KRZ BHs reported in Figs.~\ref{fig:grrt} and
    \ref{fig:cross-cut}. \textit{Right panel:} The same as on the
      left but for the DSSIM image-comparison metric.}
    \label{fig:nCCC}
\end{figure}

The left panel of Fig.~\ref{fig:nCCC} shows the image-comparison metric
in terms of the ``mismatch'' $1-{\rm nCCC}$ for different beam sizes for
KRZ BHs with an internal resolution of $\sim 0.45\,\mu{\rm as}$, hence
much smaller than that of the future projects considered. The various
colour-shaded regions show the variations of the mismatch between a Kerr
BH and KRZ BH with properties set by the given colormap. Clearly, in all
cases the mismatches decrease monotonically with the beam size or,
equivalently, with a lower angular resolution. At the same time, it is
easy to appreciate what angular resolution is needed to discriminate two
BHs with a given mismatch. As representative examples, we report with
vertical lines the present EHT resolutions and the expected ones for
ngEHT and BHEX. We should remark that a mismatch in the images of a given
amount does not translate into a comparable difference in the
spacetimes. This is because the mismatch depends sensitively on the
experiment and the latter can have a very small image mismatch between
two spacetimes that differ significantly. This behaviour is shown clearly
in the left panel of Fig.~\ref{fig:nCCC}, where the mismatch tends to
zero for very poor resolutions (i.e., very large beam sizes). Stated
differently, an inaccurate experiment will not be able to tell apart two
black-hole spacetimes even if they differ significantly. On the other
hand, increasingly accurate experiments will be able to distinguish more
easily the differences in the spacetimes. Stated differently, for two BH
images to be considered different it is sufficient that they are so as
measured in terms of the ${\rm nCCC}$ mismatch. In a similar manner, the
right panel of Fig.~\ref{fig:nCCC} reports the mismatch computed in terms
of the structural dissimilarity and can be therefore considered as the
equivalent of Fig.~\ref{fig:nCCC}, but for the DSSIM image-comparison
metric. Note that, in this case, the mismatch is overall smaller than
that reported for the ${\rm nCCC}$ metric (i.e., ${\rm DSSIM}/{\rm nCCC}
\sim 1/3$). Overall, the results shown in Fig.~\ref{fig:nCCC} highlight
that, for instance, ngEHT (BHEX) will be able to distinguish BHs with an
image mismatch $1-{\rm nCCC} \gtrsim 4.8\%$ ($1-{\rm nCCC} \gtrsim
8.0\%$); reported with different symbols are the mismatches corresponding
to the three KRZ BHs reported in Figs.~\ref{fig:grrt} and
\ref{fig:cross-cut} (see Supplementary Information Section C
  and Supplementary Figs. 5 and 6 for measurements of spacetime
  deviations not based on image-comparison metrics).

As a caveat we remark that, ideally, the image-comparison metric should
be computed making use of synthetic images produced exploiting the
visibility amplitudes of the different VLBI arrays in the ngEHT and BHEX
projects. In practice, however, since these visibility amplitudes are not
known yet under realistic conditions it is reasonable to assume that they
will be similar to those measured for the EHT and differ from the latter
only in terms of beam size. Obviously, this assumption will need to be
verified once the effective visibilities will become available. Finally,
as an interesting comparison with a specific BH in an alternative theory
of gravity, namely, a BH in the Einstein-Maxwell-dilaton-axion gravity
(dilaton BH, hereafter), a red line shows the mismatch in this case,
highlighting that the results discussed for generic KRZ BHs, applies also
when considering BHs in specific theories.

\section{Discussion}
\label{sec12}

In 2018, before the EHT Collaboration had revealed the first image of
M87$^*$, some of us explored the ``current'' (at that time) ability to test
theories of gravity with BH shadows~\cite{Mizuno:2018lxz}. The conclusion
drawn then when comparing a Kerr BH with a dilaton BH was that
distinguishing the two BHs was extremely challenging given the nominal
angular resolution of the EHT in 2018. Seven years after that result, and
with two supermassive BHs imaged by the EHT at increasingly higher
resolutions, we explore a similar question and explore the ``future''
ability of planned projects such as ngEHT and BHEX to distinguish
different BH spacetimes.

Rather than focusing on specific BHs, which would make our analysis
inevitably limited, we have adopted an accurate and yet generic
parameterisation of axisymmetric BH spacetimes, the KRZ parameterisation,
and considered some extreme corner-cases in the space of possible
parameters. After performing 3D GRMHD simulations of magnetically
arrested accretion flows (MAD)~\cite{Narayan:2003by} onto three
representative KRZ BHs and analysing the corresponding images via GRRT
calculations, we have quantified the extent to which to BH images can
differ. More specifically, we have shown that while there are qualitative
and quantitative differences in the bulk MHD properties of the accretion
and consequently on the corresponding images, these differences are also
rather small.

The importance of these results is that they provide an agnostic and
therefore generic confirmation of the ability of future experiments such
as ngEHT and BHEX to distinguish, and with extreme precision, different
BH spacetimes. At the same time, these results also stress that even
longer baseline space-VLBI observations or additional
  information -- coming either from time variability, polarization maps,
  spectral-index maps, or rotation-measure maps --  will be needed for
excluding those spacetimes that differ from the Kerr solution only
minimally.

Before concluding, a few caveats should be added. First, while our
simulations have focused on the most popular MAD accretion mode and have
adopted standard values for the adiabatic index of the equation of state
or for the electron energy distribution, differences in the simulations
and images could be introduced when considering accretion models that
differ in the accreted magnetic flux (e.g., a SANE accretion mode) or
different electron energy distributions (see also Supplementary
  Information Section D and Supplementary Figs. 7 and 8.)  While we
expect that very similar conclusions will be drawn also in this larger
space of parameters, this expectation should be confirmed via
simulations.  Second, we have modelled the expected visibilities from
ngEHT and BHEX mostly as a change in the effective beam-size. While this
is a reasonable first approximation~\cite{Fromm:2021flr}, a more precise
estimate of the mismatch will be possible when the final visibilities of
these future projects are established. Finally, the use of the KRZ metric
has the important advantage of being potentially able to describe any BH
solution known and unknown. At the same time, it has the disadvantage
that it does not provide any information on the Lagrangian and hence on
the actual theory of gravity behind the BH solution. Lacking this
information, we can only assume that the dynamics of matter and radiation
are those dictated by the KRZ metric and not by any other effective
metric, as is the case, for instance, in BHs within a nonlinear
electrodynamics description. Fortunately, preliminary and simplified
analyses in this direction have shown that the differences between
effective and background metric lead only to rather small variations in
the BH images~\cite{Kumar2024}. We plan to explore these variations via
full GRMHD simulations in future work.

\section{Methods}
\label{methods}

In what follows we present details on the methods employed for the GRMHD and
GRRT simulations, as well as the image comparison.

\subsection{Details on the GRMHD simulations}

The GRMHD simulations were performed with the help of
\texttt{BHAC}~\citep{Porth:2016rfi, Olivares:2019dsc} and the use of the
horizon-penetrating form of the KRZ metric (see Supplementary Information section A for details) and studied the impacts of the
deviation parameters in highly magnetized (MAD) flows through 3D GRMHD
simulations. To have very similar initial conditions for all
  the simulations, we fix the inner edge of the torus at $r_{\rm in}=
  20.0\,r_{g}$ and use the specific angular momentum (which is not
  constant) to tune the rest-mass density and the location of its maximum
  such that all tori have the same total rest-mass~\cite[see][for
    details]{Uniyal:2024sdv}; here $r_{g} := GM/c^2$ and $M$ is the
gravitational radius mass of the BH. We consider an ideal-fluid equation
of state with adiabatic index $\Gamma_{g} = 4/3$~\cite{Rezzolla:2013dea},
as standard in these simulations. The initial magnetic field is seeded
with a single dipolar loop $A_\phi= \left(\rho/\rho_{max}\right) \left( r
\sin\theta / r_{in} \right) \exp\left(-r/400\right) - 0.01$, such that
the radial distribution of the magnetic field supplies sufficient
magnetic flux onto the BH to produce the MAD state~\cite{Narayan:2003by,
  tchekhovskoy2011efficient}. Additionally, we used a fixed mesh
refinement approach with an effective resolution $384 \times 192 \times
192$ (over three levels), where the highest resolution is concentrated in
$|\theta-\pi/2|\le \pi/4$ and $r<100\,M$.

In order to avoid the very low-density fluid region, we fixed the floor
values of rest-mass density, $\rho_{\rm fl} = 10^{-4} r^{-2}$ and the gas
pressure, $p_{\rm fl}=(10^{-6}/3) r^{-2\Gamma_{g}}$. Similarly, the
ceiling value for the high magnetisation region is fixed by $\sigma_{\rm
  max}=100$. We also set the floor and ceiling to the electron pressure
for the electron entropy such that if the pressure is less than $1\%$ of
the floor value of the gas pressure $p_{\rm fl}$, we set $p_{e} = 0.01
p_{\rm fl}$. Similarly, if electron pressure becomes larger than the gas
pressure, we set $p_{e} = 0.99\, p_{\rm fl}$. It is important to stress
that these choices are the standard ones and have been validated in a
very large number of simulations~\cite{EventHorizonTelescope:2019pcy}.
The jet power $P_{\rm jet}$ employed in the main text as an important
measure to distinguish different BH spacetimes is calculated by taking
the integration over the $2$-sphere at $r=50 \, r_g$
\citep[e.g.,][]{Nathanail-etal2020, Dihingia-etal2021}
\begin{equation}
  \begin{aligned}
    P_{\rm jet}:=\int_0^{2\pi}\int^\pi_0 (-T^r_t -\rho u^r)\sqrt{-g}d\theta
    d\phi\,,
    \label{pjet}
  \end{aligned}
\end{equation}
where the integrand in the above equation is set to zero if $\sigma \leq
1$ over the integrating surface. Once again, this represent the standard
procedure to compute $P_{\rm jet}$.

\subsection{Details on the GRRT simulations}

We perform GRRT calculations for the 3D MAD accretion scenarios with the
help of the GRRT code \texttt{RAPTOR}~\cite{Bronzwaer:2018lde,
  Bronzwaer:2020kle}. The target source is taken to be Sgr~A$^*$, thus
having mass $M=4.14 \times 10^6$ \(\textup{M}_\odot\) and distance
$8.127\,{\rm kpc}$~\cite{EventHorizonTelescope:2022wkp}; however, very
similar results would be obtained if we considered M87$^*$ The pixel
resolution in the image plane is $500 \times 500$ and the image field of
view is set to be $0.25\, {\rm mas}^2$ $(40 \, M^2)$. As customary in
supermassive BH imaging, we only consider synchrotron radiation with a
thermal electron distribution function for all the cases. The electron
temperature is calculated using an $R\!-\!\beta$ parameterized
prescription, which uses the ion-to-electron temperature
ratio~\cite{moscibrodzka2009radiative}
\begin{equation}
\frac{T_i}{T_e}=\frac{\left(R_{\rm low}+R_{\rm
    high}\beta^2\right)}{\left(1 + \beta^2\right)}\,.
\label{eq:Ti_Te}
\end{equation}
Here, $\beta=p_g/p_m$ is the plasma beta, i.e., the ratio of gas pressure
to the magnetic pressure. We fixed $R_{\rm low}=1$ and $R_{\rm high} =
160$, consistent with the current EHT
observations~\cite{EventHorizonTelescope:2022urf}, and a standard choice
in simulations of this type. However, our results do not depend
appreciably on the choice of $R_{\rm high}$, as discussed in detail in
Supplementary Information section B. Finally, as a useful reference, the
inclination angle is fixed at at $i=30^\circ$; this is consistent with
the EHT and GRAVITY constraints~\cite{EventHorizonTelescope:2022urf,
  GRAVITY:2020hwn}, but different values of $i$ would lead to very
similar quantitative results (see Supplementary Information
  Section D and Supplementary Figs. 9 and 10).

\subsection{Image comparisons}
\label{sec:im_comp}

For a quantitative image comparison, we follow Refs.~~\cite{Wang2004,
  Mizuno:2018lxz} and adopt as metric of the differences in the images
the normalized cross-correlation coefficient (nCCC). This coefficient can
be considered as a measure of the global ``overlap'' between two images,
so that ${\rm nCCC}=1$ when the two images are identical and ${\rm
  nCCC}=0$ when two images are entirely distinct. The ``mismatch'', i.e.,
$1-{\rm nCCC}$, is then computed as
\begin{equation}
  1- {\rm nCCC}(I,K):= 1 - \frac{1}{N} \sum_i \frac{(I_i - \mu_I) (K_i -
    \mu_K)}{\sigma_I \sigma_K}\,,
  \label{nccc_eq}
\end{equation}
where $\mu_I$ and $\mu_K$ are the mean pixel value in the two images $I$
and $K$, $\sigma_I$ and $\sigma_K$ are the standard deviations of the
pixel values for two images. The sum is done over all $N$ pixels in both
images.

An alternative measure can be made with the so-called ``structured
dissimilarity'' (DSSIM) index, which is computed by first calculating
``structural similarity (SSIM)'' as
\begin{equation}
    \begin{aligned}
  {\rm SSIM}(I,K):= \mathcal{S}(I,K)\, \mathcal{C} (I,K)\,,
\label{ssim_eq}
\end{aligned}
\end{equation}
where $I$ and $K$ refer to the image pairs that is being compared, while
$\mathcal{S}(I,K)$, and $\mathcal{C} (I,K)$ are the ``structure'' and
``contrast'' dissimilarities defined respectively as
\begin{align}
  {\mathcal S}(I,K) &:= \frac{2 \sigma_I \sigma_K}{\sigma^2_I +
    \sigma^2_K} \,,\\ {\mathcal C}(I,K) &:= \frac{\sigma_{IK}}{\sigma_I
    \sigma_K} \,.
    \label{ssim_eq_2}
\end{align}
In the definitions above, we have used the following quantities as
shorthands
\begin{eqnarray}
  \mu_I&:=&\sum_i \frac{I_i}{N}\,,\\
  \sigma^2_I&:=& \frac{\sum \limits^N_{j=1} (I_j - \mu_j)^2}{(N-1)}\,,\\
  \sigma_{IK}&:=& \frac{\sum \limits^N_{j=1} (I_j - \mu_I) (K_j - \mu_K)}{(N-1)}\,,
\end{eqnarray}
where $I_i$ is the intensity of the $i$-th pixel of image $I$. Using now
Eqs.~\eqref{ssim_eq_2} and \eqref{ssim_eq_2}, the DSSIM can be written as
\begin{equation}
    {\rm SSIM}(I,K) = \frac{2 \sigma_{IK}}{\sigma^2_I + \sigma^2_K} \,,
    \label{ssim_eq_3}
\end{equation}
and we can again compute the mismatch as ${\rm DSSIM}:=1/|\rm SSIM|-1$,
so that two images are identical if ${\rm DSSIM} = 0$ and they differ
maximally if $\rm DSSIM = 1$.

 \backmatter \bmhead{Data Availability} The simulation data and
  analysis scripts used in this work are available upon reasonable
  request.

\bmhead{Code Availability} The publicly released version of the GRMHD code 
BHAC and GRRT code RAPTOR can be found at https://bhac.science and 
https://github.com/jordydavelaar/raptor.

\backmatter \bmhead{Acknowledgements} We thank K. Moriyama and Y.  Ma for
discussions and C. Fromm, and Z. Younsi for insightful comments and
useful suggestions. This research is supported by the National Key R\&D
Program of China (Grant No.\,2023YFE0101200), the National Natural
Science Foundation of China (Grant No.\,12273022), the Research Fund for
Excellent International PhD Students (grant No. W2442004) and the
Shanghai Municipality orientation program of Basic Research for
International Scientists (Grant No.\,22JC1410600), the ERC Advanced Grant
``JETSET: Launching, propagation and emission of relativistic jets from
binary mergers and across mass scales'' (Grant No. 884631), and the
European Horizon Europe staff exchange (SE) programme
HORIZON-MSCA2021-SE-01 Grant No. NewFunFiCO-101086251. LR acknowledges
the Walter Greiner Gesellschaft zur F\"orderung der physikalischen
Grundlagenforschung e.V. through the Carl W. Fueck Laureatus Chair and
the hospitality at CERN, where part of this research was carried out. The
simulations were performed on TDLI-Astro cluster and Siyuan Mark-I at
Shanghai Jiao Tong University.

\bmhead{Author Contributions} AU performed the GRMHD and GRRT simulations
and wrote the initial draft. ID, YM and LR provided insight into the
scientific interpretation of the results. LR initiated and closely
supervised the project, and wrote the manuscript.  All authors discussed
the results and commented on all versions of the
manuscript.

\bmhead{Competing interests} The authors declare no competing interests.





\newpage

\title{\Huge Supplementary Materials}
\maketitle
\vspace{2em}
\section*{Section A: Horizon-Penetrating form of KRZ}
\label{HP-KRZ}

In order to perform long-term stable simulations of BHs, a
horizon-penetrating form of the metric is required because in these
coordinates, all metric coefficients remain regular at the event
horizon. As a result, we consider the horizon-penetrating form of the
parameterized KRZ metric (HP-KRZ)~\cite{Ma:2024kbu} exploiting a special
class of the KRZ solutions that allows for a separable form of the
Hamilton-Jacobi equations~\cite{Konoplya2018}. Hence, the line element of
the HP-KRZ metric reads
\begin{align}
 \label{eq:HP_KRZ_metric}
 ds^2 = &-\left(1 - \frac{R_M}{\Sigma r}\right) d t^2 + 2 \frac{R_M}
 {\Sigma r} R_B d t d r - 2 \frac{R_M}{\Sigma r} a \sin^2{\theta} d t d
 \phi \nonumber \\ &+ \left(1 + \frac{R_M}{\Sigma r}\right) R_B^2 d r^2 -
 2 \left(1 + \frac{R_M}{\Sigma r}\right) R_B a \sin^2{\theta} d r d
 \phi + \Sigma r^2 d \theta^2 + K^2 r^2 \sin^2{\theta} d \phi^2 \,,
\end{align}
where $\Sigma:=1+{a^2 \cos^2{\theta}}/{r^2}$ and
\begin{subequations}
\begin{eqnarray}
R_B&:=&1+b_{00}(1-x)+\dfrac{b_{01}(1-x)^2}{1+\dfrac{b_{02}x}
  {1+\dfrac{b_{03}x}{1+\ldots}}}\,,\\ R_M&:=&r_0\Biggr(1+\frac{a^2}{r_0^2}
(1-x)^2+\epsilon_0x \\\nonumber&&-(a_{00}-\epsilon_0)(1-x)x-
\dfrac{a_{01}(1-x)^2x}{1+\dfrac{a_{02}x}{1+\dfrac{a_{03}x}{1+\ldots}}}\Biggr)\,,\\
K^2&:=&\Sigma
+ \left( 1 + \frac{R_M}{\Sigma r} \right) \frac{a^2
  \sin^2{\theta}}{r^2}\,.
\end{eqnarray}
\end{subequations}
Here $a$ is the dimensionless spin of the BH, $x := 1-{r_0}/{r},$ and
$r_0$ is the horizon radius in the equatorial plane. The coefficients
$a_{00}$ and $b_{00}$ are the asymptotic parameters, $\epsilon_0$
provides an additional freedom in setting the location of the event
horizon, whereas the coefficients $b_{01},b_{02},b_{03},\ldots$ and
$a_{01},a_{02},a_{03},\ldots$ describe the near-horizon geometry.

Exploiting the rapid convergence of the KRZ expansion that allows to use
only the first-order coefficients to reach an accurate description of the
spacetime~\cite{Kocherlakota:2020kyu}, but also to keep our treatment to
a limited set of coefficients, we here focus only on the coefficients
that affect the angular frequency $\Omega=g_{t \phi}/g_{tt}$. As a
result, we restrict our analysis to the two parameters $a_0 := a_{00}$
and $a_1 := a_{01}$ and fix the coefficient $\epsilon_0$ to place the
event horizon as in the case of a Kerr BH, i.e., $\epsilon_0 =
(2M-r_0)/r_0$. 

In this case, the above expressions reduce to
\begin{subequations}
\begin{eqnarray}
  R_B&=&1.0\,,\\ R_M&=&r_0\Biggr(1+\frac{a^2}{r_0^2}(1-x)^2 +
  \left(\frac{2-r_0}{r_0}\right) x\\\nonumber&&- \left(
  a_{0}-\frac{2-r_0}{r_0} \right)(1-x)x-a_{1}(1-x)^2x\Biggr)\,.
\end{eqnarray}
\end{subequations}

In Supplementary Fig.~\ref{metric}, we show the $g_{t\phi}$ component of
the HP-KRZ metric in the equatorial plane $(\theta=\pi/2)$ for BH spin
$a=0.9375$. Clearly, all the metric functions converge to the Kerr value
at the horizon and tend to unity at spatial infinity. At the same time,
in the vicinity of the event horizon, the KRZ BHs will have metric
functions that differ significantly from the Kerr solution, with the KRZ
BH given by $a_0=1.0=a_1$ showing the most significant deviations, and
which can be as large as $25\%$. Overall, our analysis has considered
three KRZ BHs ($[a_0=1.0,\, a_1=0.0]$, $[a_0=0.0,\, a_1=1.0]$, and
$[a_0=1.0,\, a_1=1.0]$) that can be thought to represent the corner cases
of the considered parameterisation and where the parameter ranges are
consistent with the mathematically allowed
limits~\cite{Kocherlakota:2020kyu, Kocherlakota:2022jnz}.

\begin{figure}
    \centering
  \includegraphics[width=0.5\textwidth]{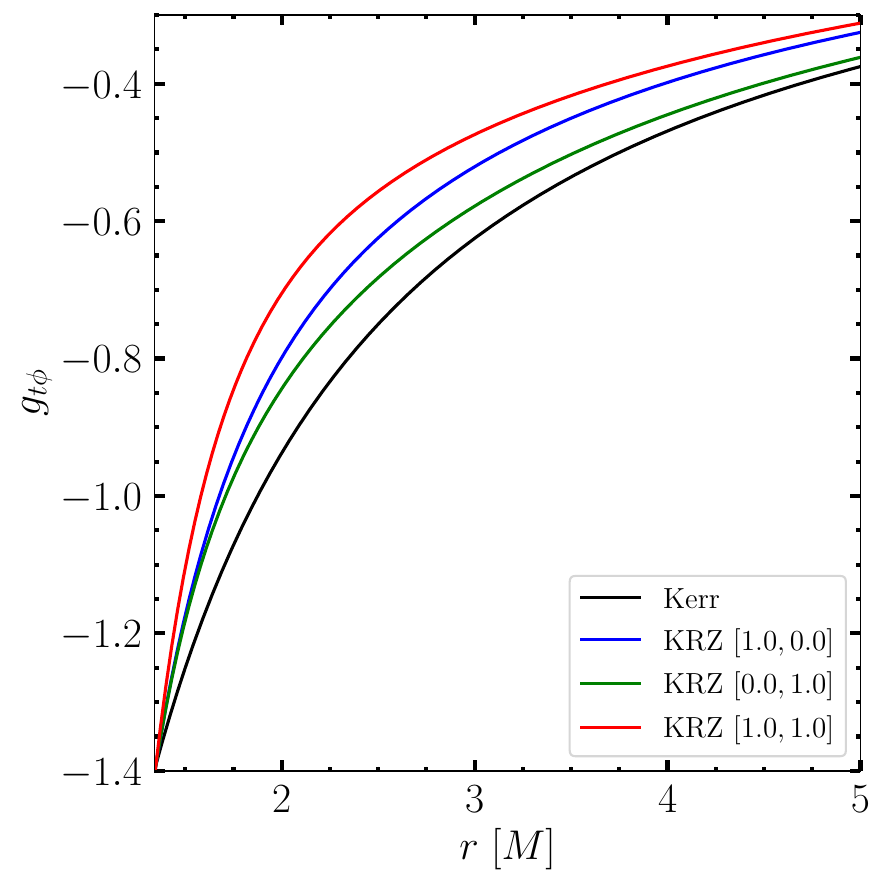}
    \caption{\textbf{Frame dragging metric component: Kerr vs. KRZ BHs.}
      $g_{t\phi}$ component of the metric for Kerr and deformed
      parameters for a BH spin $a=0.9375$ in the equatorial plane
      $(\theta=\pi/2)$.}
    \label{metric}
\end{figure}

\section*{Section B: Selected output of the GRMHD simulations}
\label{sec:GRMHD_out}

In what follows, we present a selection of the output of the GRMHD
simulations that complements the information provided in the Main text
and in the Methods section.

\begin{figure}
  \centering
  \includegraphics[width=0.75\textwidth]{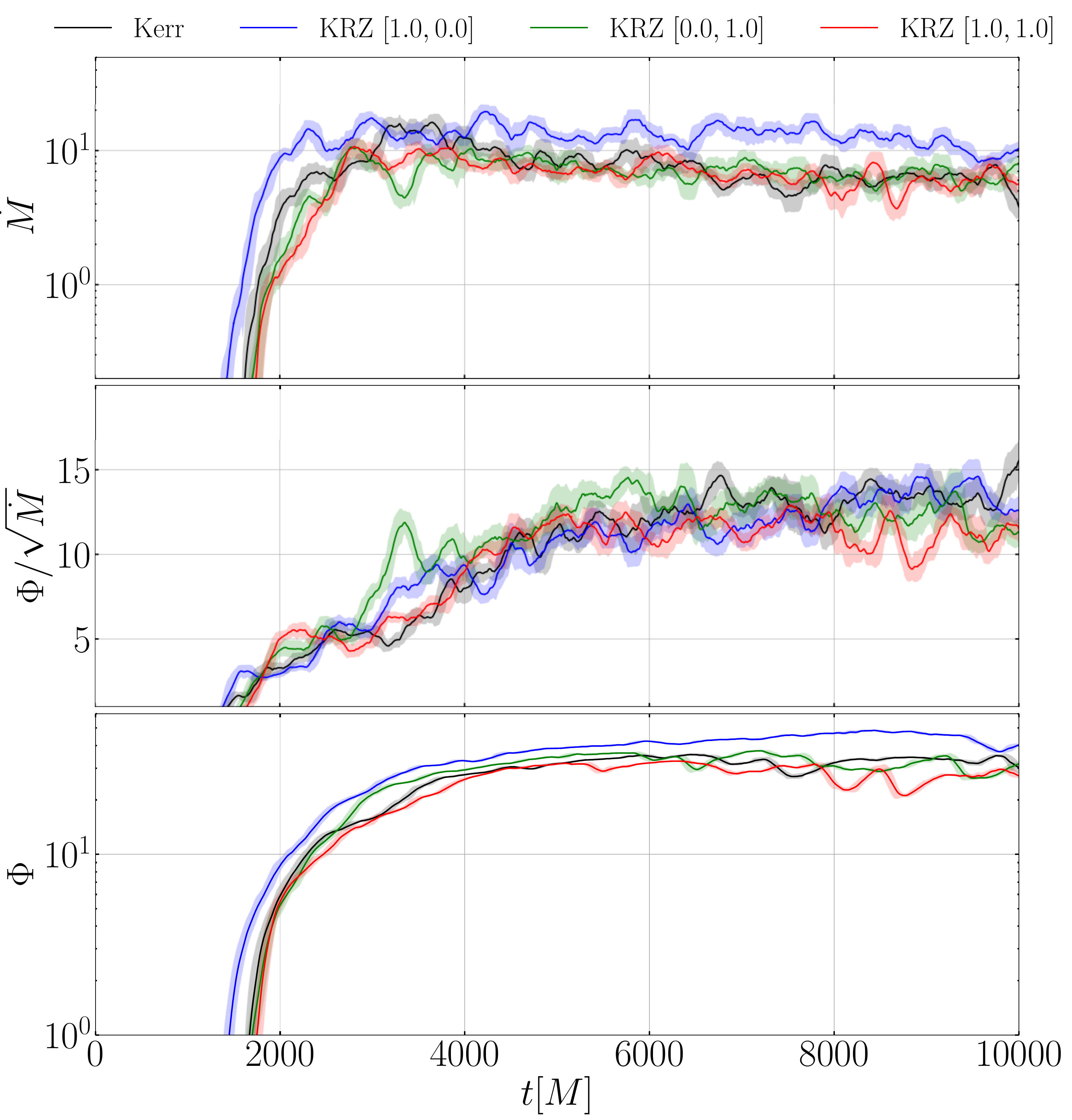}
    \caption{\textbf{Running average and standard-deviation bands of
        horizon quantities.} The running average and transparent standard
      deviation variation bands of volume integrated mass accretion rate
      ($\dot{M}$, upper panel), normalized magnetic flux ($\phi /
      \sqrt{\dot{M}}$, middle panel), and magnetic flux ($\phi$, lower
      panel) at the horizon of the BH for the fixed spin $a=0.9375$.}
    \label{flux_3d}
\end{figure}

We start with Supplementary Fig.~\ref{flux_3d} that reports a number of
quantities normally monitored in GRMHD simulations of accretion flows
onto BH, namely, the volume-integrated mass-accretion rate $(\dot{M})$,
the normalized magnetic flux $(\phi / \sqrt{\dot{M}})$, and the magnetic
flux $(\phi)$ calculated at the horizon (the solid lines refer to the
average values and the shading shows one standard-deviation
variation). Note that all quantities have reached a steady-state
evolution by $t=10,\!000\,M$ and that the normalized flux reaches values
$\phi/\sqrt{\dot{M}} \simeq 15$, as defined for a MAD accretion
mode~\citep{Tchekhovskoy:2012bg}. Note also that the behaviour is very
similar across all BHs considered, remarking again the difficulties in
distinguishing these BHs using these quantities.

Similarly, shown in Supplementary Fig.~\ref{fig:poltotor} with a colormap
is the spatial distribution of the ratio of the poloidal-to-toroidal
magnetic-field components for a Kerr BH (left panel) and for KRZ BHs (all
the other panels). The information in this figure, which again refers to
data that is time-averaged in the window $t=8,\!000-10,\!000 \, M$,
complements the one already discussed in Fig. 2 when presenting the
$\sigma=1$ contours of the jet and of the disc in the various
spacetimes. In particular, besides noting that the Kerr BH is the one
with the most collimated jet and most compact disc, it is possible to
appreciate that the poloidal magnetic field is comparable to the toroidal
one for a Kerr BH. On the other hand, KRZ BHs have poloidal magnetic
fields that are stronger than the toroidal ones and up to about one order
of magnitude. These findings, together with differences found in the jet
power, provide evidence that additional information on the spacetime
properties can probably be found in the electromagnetic properties of the
accretion process, thus opening the way to additional input to break the
degeneracy in the observed horizon-scale images. We plan to explore these
signatures in a more systematic future study.

\begin{figure}
  \centering
  \includegraphics[width=0.75\textwidth]{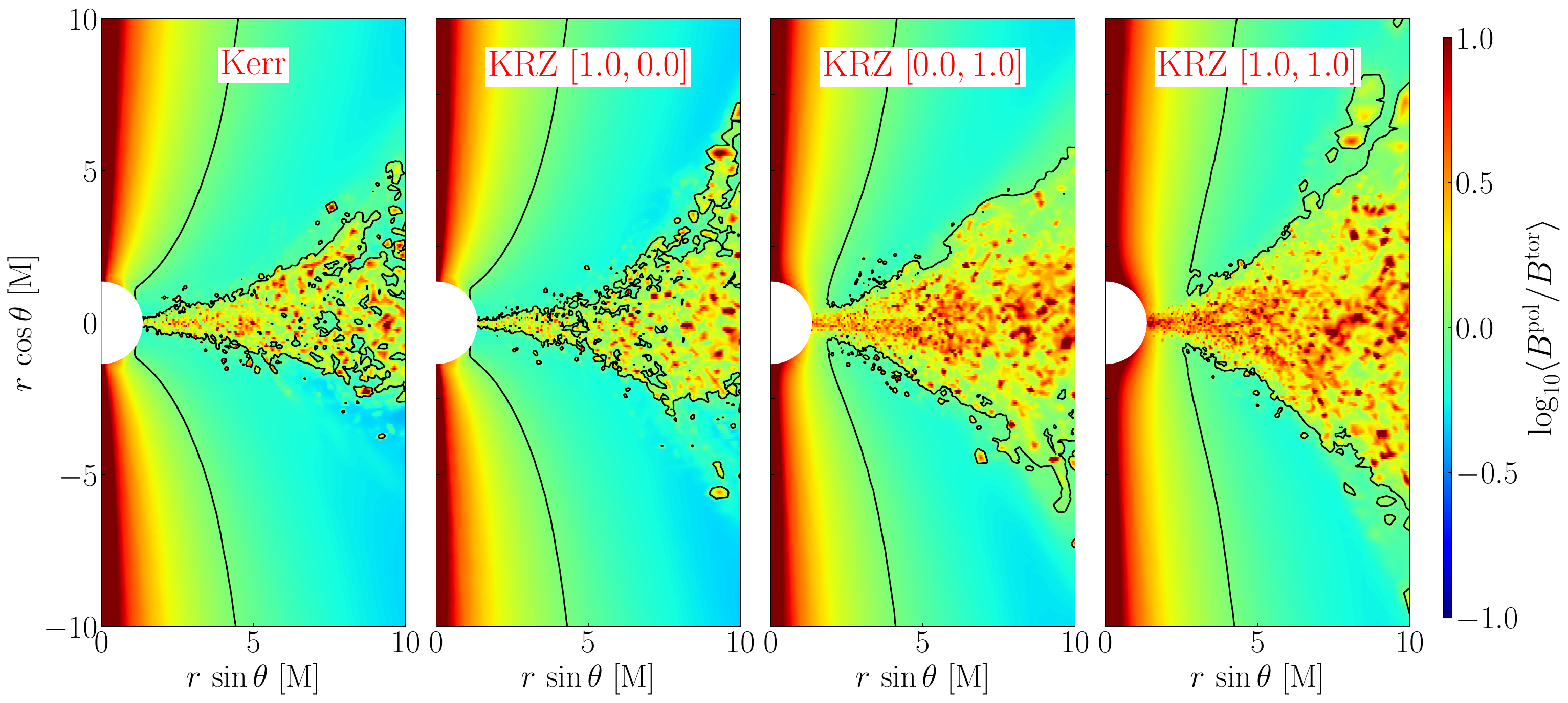}
  \caption{\textbf{Comparision of poloidal-to-toroidal magnetic-field
      ratio.} Spatial distribution of the ratio of the
    poloidal-to-toroidal magnetic-field components for a Kerr BH (left
    panel) and for KRZ BHs (all the other panels). Also in this case, the
    data is time-averaged in the window $t=8,\!000-10,\!000 \, M$. The
    information in this figure complements the one already shown in
    Figs. 1 and 2.}
  \label{fig:poltotor}
\end{figure}

Finally, another quantity customarily monitored in these simulations is
the so-called ``MRI quality factor'', namely, the number of cells
available to resolve the fastest growing MRI mode. A convenient way to
compute this is to compare the wavelength of the fastest growing MRI mode
$\lambda^{(\alpha)}$ in the tetrad basis of the fluid frame
$e_\mu^{(\alpha)}$
\begin{align}
\lambda^{(\alpha)} := \frac{2\pi}{\sqrt{(\rho h + b^2)\Omega}}b^\mu
\boldsymbol{e}_\mu^{(\alpha)}\,,
\end{align}
where $\Omega=u^\phi/u^t$ is the angular velocity, with the grid
resolution as seen in the orthonormal fluid-frame $\Delta x^{(\alpha)} =
\Delta x^\mu \boldsymbol{e}_\mu^{(\alpha)}$. In this way, it is possible
to define the quality factor as $Q^{(\alpha)} :=\lambda^{(\alpha)}/\Delta
x^{(\alpha)}$~\citep{Takahashi:2007dk, Siegel:2013nrw, Porth:2016rfi}
where and report it at $t=10,\!000 \, M$ for the four spacetimes
considered here. The quality factors in the radial and polar directions
-- $Q^{(r)}$ (upper panels) and $Q^{(\theta)}$ (lower panels) -- are
shown in Supplementary Fig.~\ref{Q-factor} and indicate that the MRI is
resolved since $Q^{(\alpha)} \gtrsim 6$ in the inner regions of the
accretion flow.

\begin{figure}
  \centering
  \includegraphics[width=0.75\textwidth]{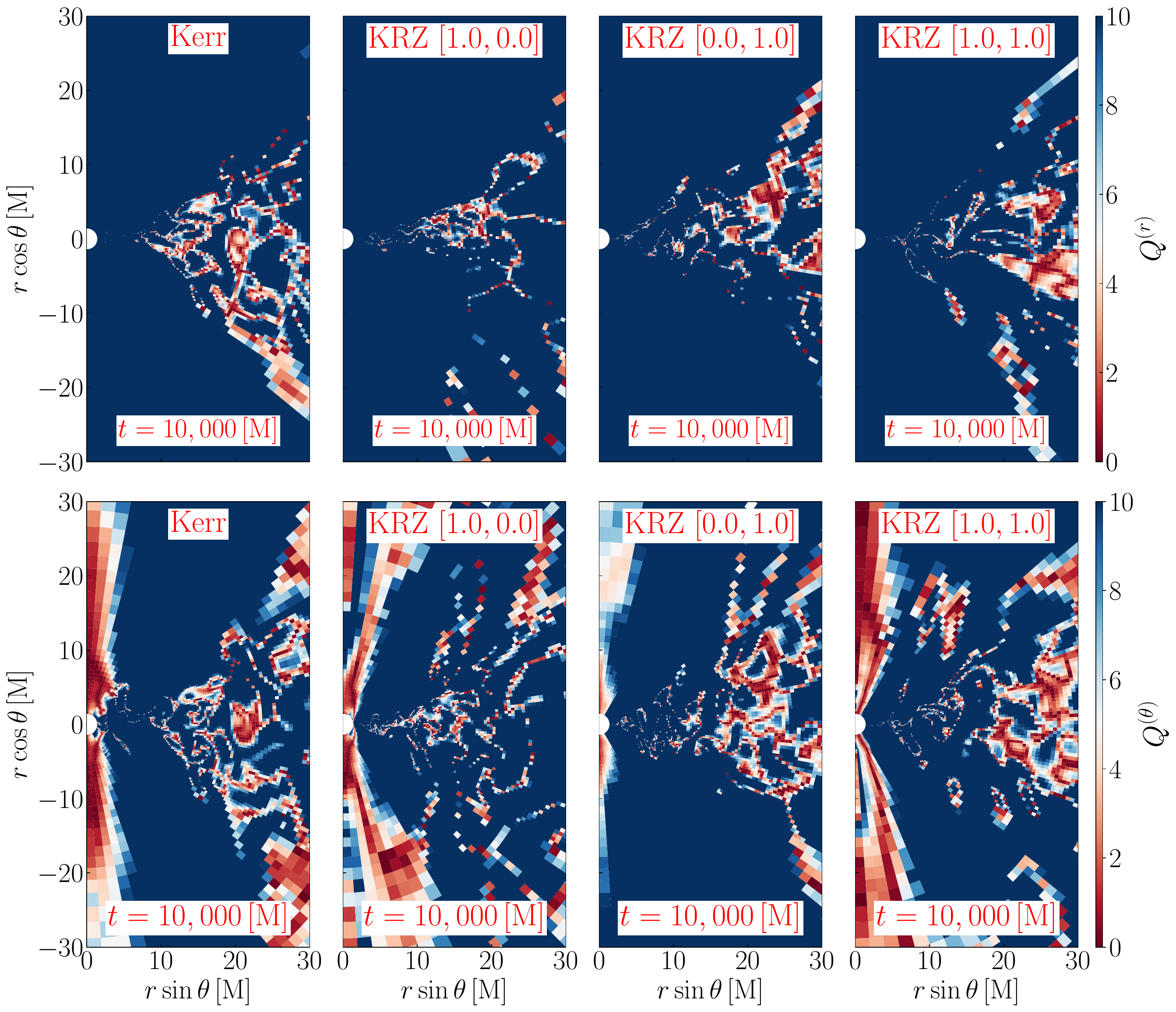}
  \caption{\textbf{MRI quality factors.} The MRI quality factor $Q^{(r)}$
    (upper panels) and $Q^{(\theta)}$ (lower panels) at $t=10,\!000 \, M$
    and for the four spacetimes considered here.}
  \label{Q-factor}
\end{figure}

\section*{Section C: Alternative metric for measuring deviations from Kerr}
\label{sec:Alternative_metric}

All of the approaches considered so far to measure deviations from the
Kerr spacetime have used image-based metrics such as nCCC or DSSIM (see
Fig. 5 and discussion in Results). We believe these metrics to be most
reliable and robust as they are intrinsically agnostic and do not rely on
the assumption that certain features, most notably, an amplified
intensity near the photon ring, are present in the actual image.
However, it is interesting and useful to consider also other approaches
to measure deviations from the Kerr spacetime based on specific features
of the images.

\begin{figure}
  \centering
  \includegraphics[width=0.5\textwidth]{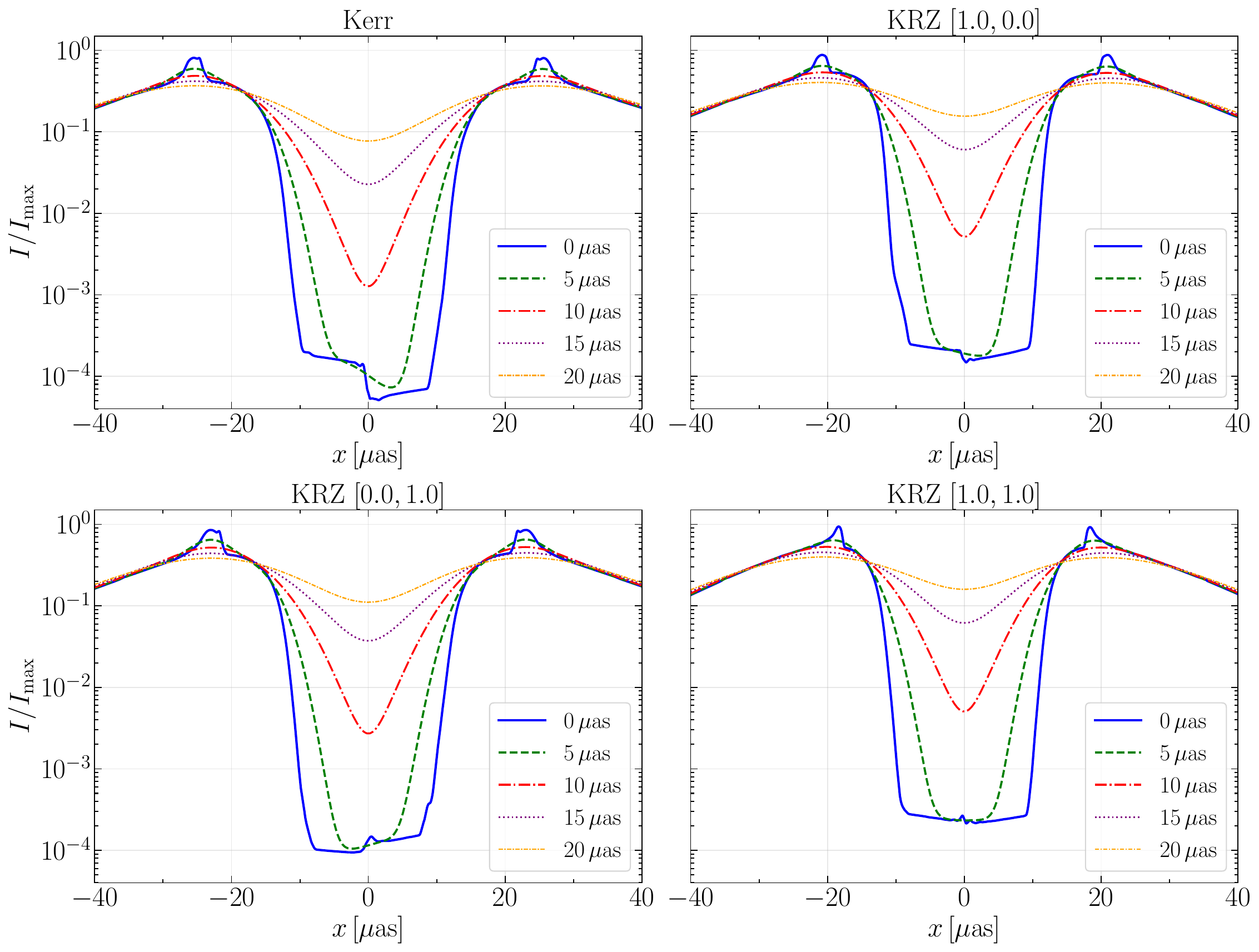}
  \caption{\textbf{Time and azimuthally-averaged intensity profiles.}
    Time and azimuthally-averaged intensity profiles (see also Fig. 4)
    for the four BHs considered here and their variation as the beam size
    increases. The profiles refer to a face-on observer, \ie with an
    inclination angle $i \simeq 0^\circ$. Note the presence of a
    so-called ``inner shadow'', whose properties are however difficult to
    measure with precision.}
    \label{fig:profile_comp}
\end{figure}

More specifically, taking as a reference Fig. 4, we here assume that the
cross sections of the image are characterised by two clear peaks in the
intensity and that these can be used to measure a ring diameter $D_p$
with a given precision $\delta D_p$, where the latter is estimated as the
width of the intensity profile at $99\%$ of its maximum value. Obviously,
the measurement of $D_p$ will converge to the exact diameter of the
photon ring in the case of infinite resolution and will degrade as the
beam size increases. Supplementary Figure~\ref{fig:profile_comp} reports
the intensity profiles for the four BHs considered here after a time and
azimuthal average intensity profiles (see also Fig. 4) and shows their
variation as the beam size increases. The profiles refer to an idealised
observation for an observer that is almost ``face-on'', \ie with an
inclination angle $i\simeq 0^\circ$ ($i=0.01^\circ)$; the uncertainty
increases for $i=30^\circ$ but not significantly. Note also the presence
of the so-called ``inner shadow''~\citep{Dokuchaeve2019, Dokuchaeve2020,
  Chael2021}, whose properties are however difficult to measure with
precision. This is because the inner shadow is not constant, but depends
on the intensity contrast and thus on the dynamic range of the measuring
network (which is difficult to estimate a-priori); furthermore, its
appearance is highly asymmetrical, thus preventing a simple measurement
of its diameter (see Supplementary Fig.~\ref{fig:profile_comp}). This
high level of uncertainty in the properties of the inner shadow limits
the precision with which deviations can be quantified, making it
challenging to draw robust conclusions based on this feature alone.
\begin{figure}
  \centering
  \includegraphics[width=0.5\textwidth]{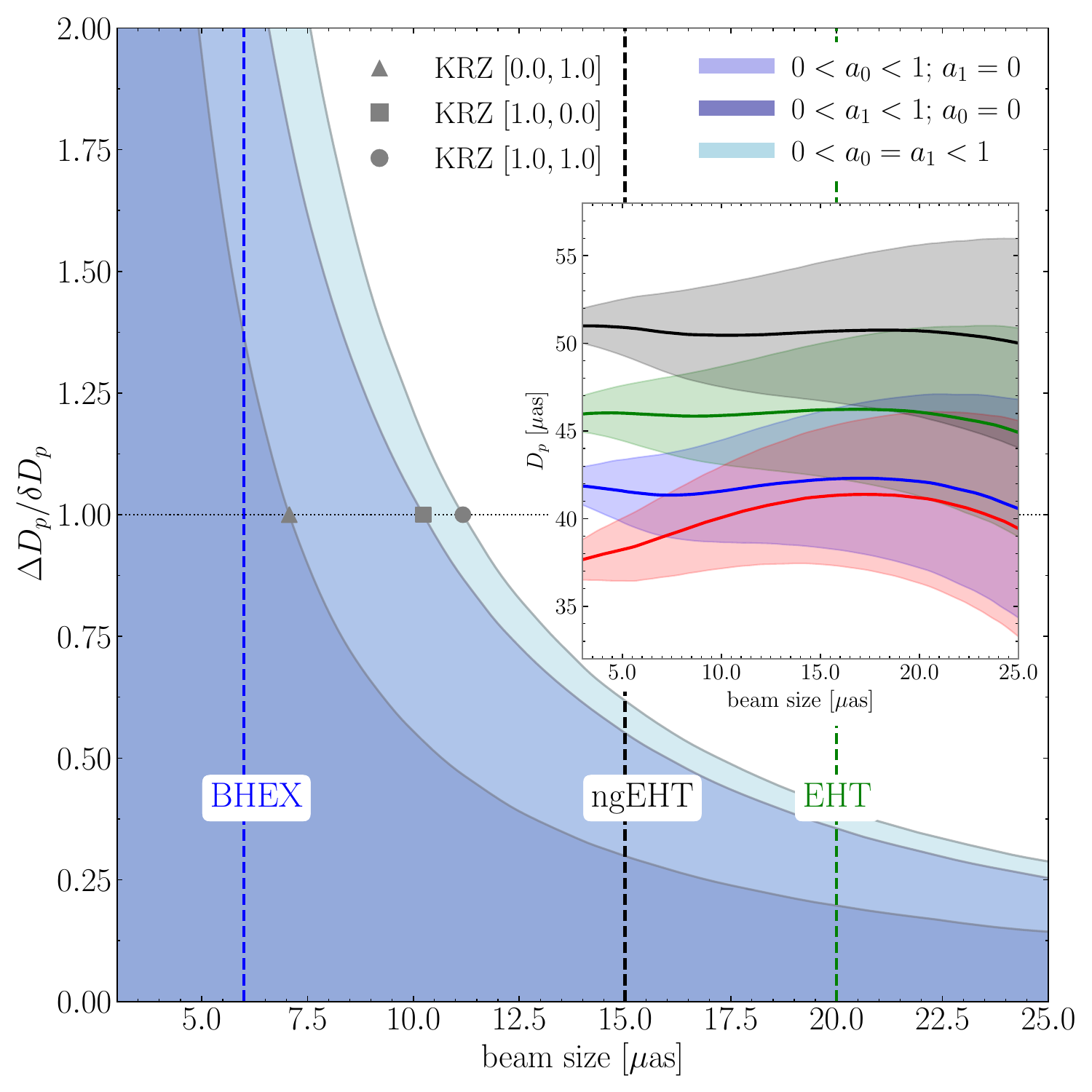}
  \caption{\textbf{Difference in ring diameter measurements for Kerr and
      KRZ BHs.} Difference in the measurement of the ring diameter
    $\Delta D_p := D^{^{\rm KRZ}}_p - D^{^{\rm Kerr}}_p$ normalised by
    its uncertainty $\delta D_p := \sqrt{[(\delta D^{^{\rm KRZ}}_p)^2 +
        (\delta D^{^{\rm Kerr}}_p)^2]/2}$ between a Kerr BH and the three
    KRZ BHs considered here. The inset displays with solid coloured lines
    the measured values of $D_p$ for the four BHs and with associated
    shaded regions the size of the uncertainty $\delta D_p$.}
    \label{fig:A_M}
\end{figure}
However, it is possible measure the deviation from the Kerr metric as
\begin{equation}
\frac{\Delta D_p}{\delta D_p} := \frac{D^{^{\rm KRZ}}_p - D^{^{\rm
      Kerr}}_p}{\sqrt{\left[\left(\delta D^{^{\rm KRZ}}_p)^2 + (\delta
      D^{^{\rm Kerr}}_p\right)^2\right]/2}}\,.
  \end{equation}
Clearly, values of ${\Delta D_p}/{\delta D_p} \gg 1$ would indicate that
the spacetimes are clearly distinguishable, while values ${\Delta
  D_p}/{\delta D_p} \ll 1$ are expected to characterise spacetimes where
the uncertainty in the measurement of the diameter dominates the measure
of the diameter. In this respect, the ratio ${\Delta D_p}/{\delta D_p}$
can be assimilated to a signal-to-noise ratio, whose ultimate value will
depend both on the BH spacetime considered and on experiment carried out
to produce the measurement.

Supplementary Figure~\ref{fig:A_M} report $\Delta D_p / \delta D_p$ for
the three KRZ BHs using a notation similar to Fig. 5 and hence as a
function of the beam size. Also reported with a horizontal dotted line is
the value $\Delta D_p / \delta D_p = 1$, so that the corresponding
intersections with the shaded areas mark the critical beam size needed to
obtain a measurement not dominated by the uncertainty. Importantly, the
inset displays with solid coloured lines the measured values of $D_p$ for
the four BHs and with associated shaded regions the size of the
uncertainty $\delta D_p$. Overall, Supplementary Fig.~\ref{fig:A_M}
nicely confirms and complements the information in Fig. 5 and
Supplementary Fig.~\ref{fig:nCCC_2} by showing that -- also when using a
completely distinct metric that is not based on global image comparison
-- distinguishing BHs with present experiments is very challenging, but
also that the expected beam sizes of future projects will be able to
tackle these challenges.

\section*{Section D: Impact of electron energy distribution and inclination angle}
\label{sec:Rhigh}

As mentioned in Methods, an important aspect of the GRRT treatment has to
do with the energy distribution of the electrons involved in the
synchrotron emissions, which we consider to be the most important one,
under the assumption that the non-thermal component is small in most of
the image morphology at $230\,{\rm GHz}$. There are several ways of
introducing a description of the energy distribution, starting from the
most sophisticated implementing turbulent and resistive
corrections~\citep{Davelaar2019, Mizuno:2021esc} or inspired by
microphysics~\cite{Meringolo2023}, the simpler ones in which one relates
the energy of the electrons to the energy of the ions as computed from
the GRMHD simulations via a simple prescription also known as the
$R\!-\!\beta$ parameterisation~\citep{Moscibrodzka:2015pda}. The latter
is the most common choice and while it is not the most realistic, it has
also been shown to to provide a reasonable
approximation~\citep{Mizuno:2021esc}. The main degree of freedom in this
parameterisation is associated with the coefficient $R_{\rm high}$ in
Eq. 2 of main text (the other coefficient $R_{\rm low}$ does not
introduce significant differences).

The GRRT simulations discussed in the main text have employed the
commonly used value of $R_{\rm high}=160$, but it is of course important
to assess how the results of the GRRT simulation and the conclusions we
draw on the ability to distinguish BH images depend on this choice. To
this scope, we have repeated the analysis also for values of $R_{\rm
  high}=40$ and $R_{\rm high}=10$, where the latter is the smallest value
that is useful to consider~\cite{EventHorizonTelescope:2019dse,
  EventHorizonTelescope:2019ths, EventHorizonTelescope:2019ggy,
  EventHorizonTelescope:2022wkp, EventHorizonTelescope:2022wok}.

\begin{figure*}
  \centering
  \includegraphics[width=0.48\textwidth]{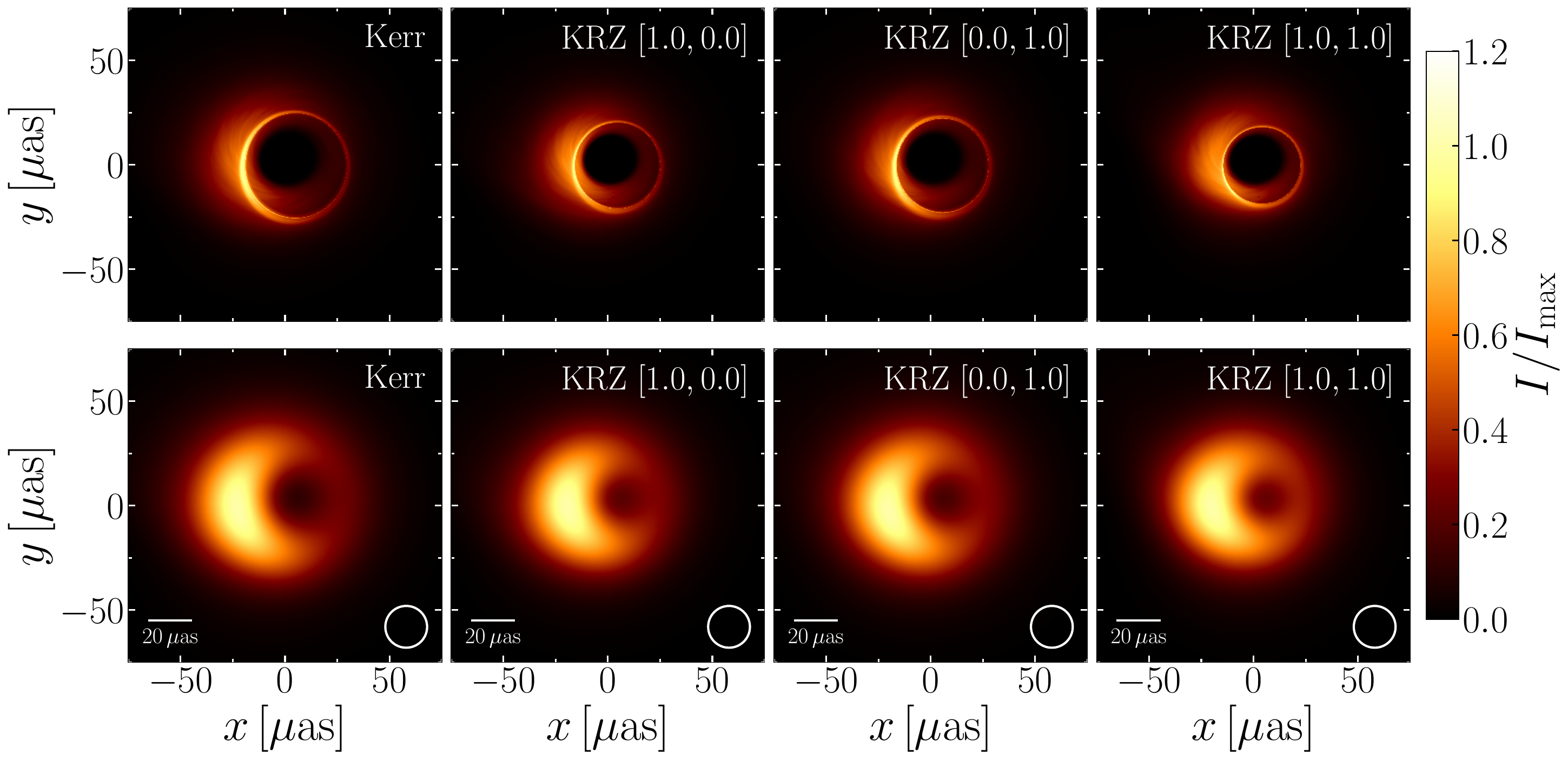}
  \hspace{0.2cm}
  \includegraphics[width=0.48\textwidth]{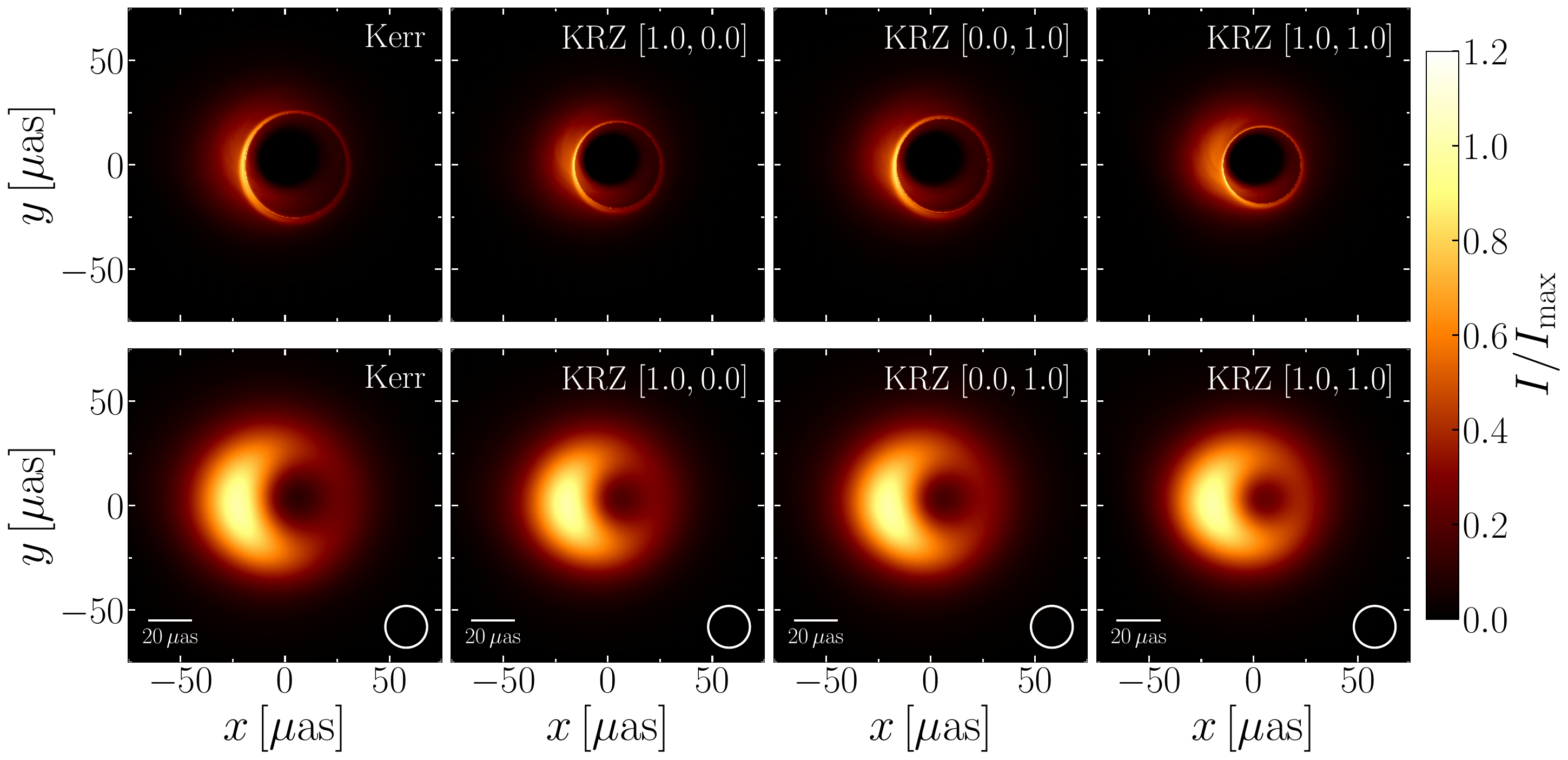}
  \caption{\textbf{Different electron-temperature parameter images.} The
    same as in Fig. 3 but for different values of the
    electron-temperature parameter, namely, $R_{\rm high}=40$ (left
    panel) and $R_{\rm high}=10$ (right panel); by contrast, $R_{\rm
      high}=160$ was employed in Fig. 3.}
    \label{grrt_2}
\end{figure*}

\begin{figure*}
  \centering
  \includegraphics[width=0.48\textwidth]{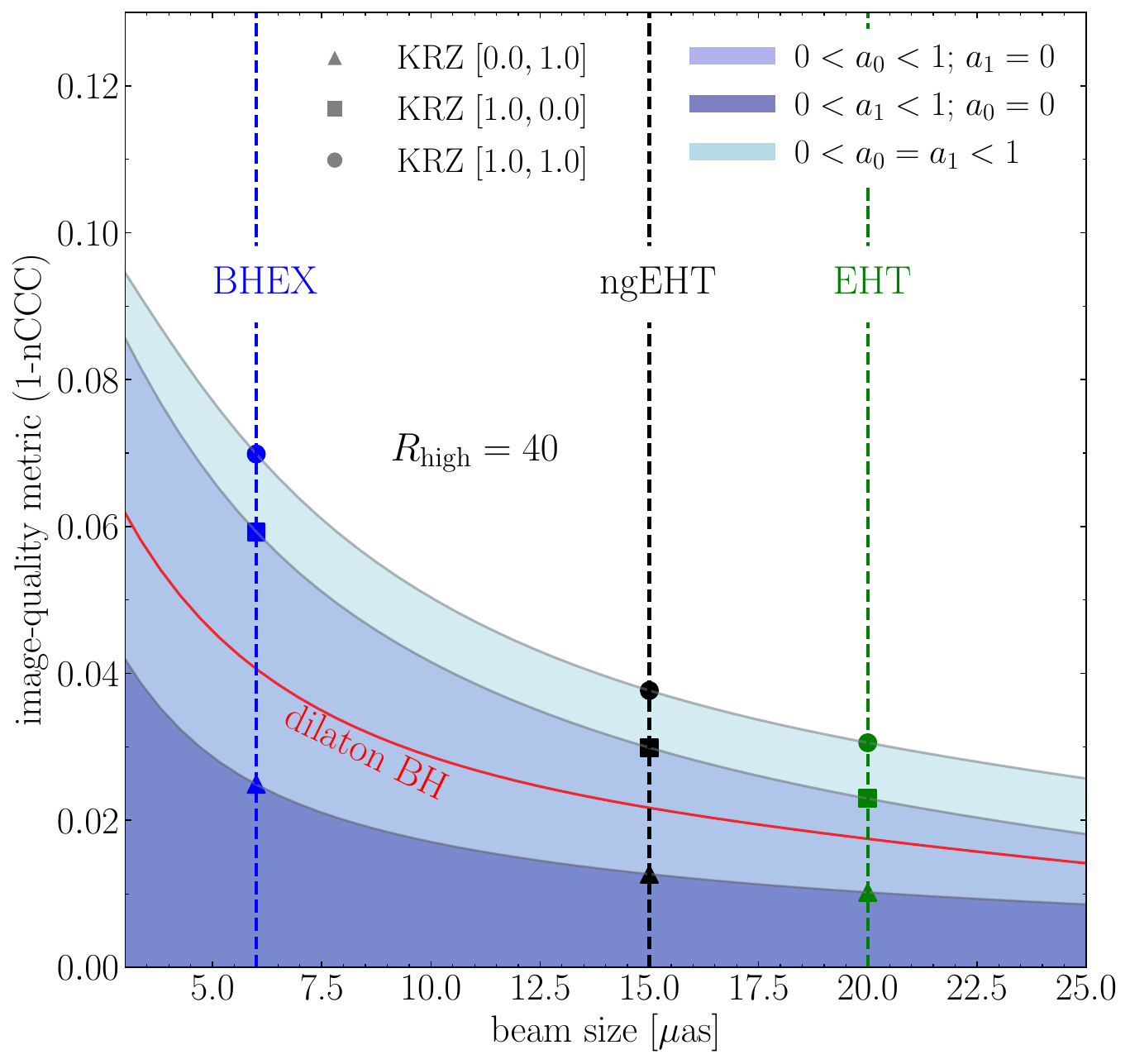}
  \hspace{0.2cm}
  \includegraphics[width=0.48\textwidth]{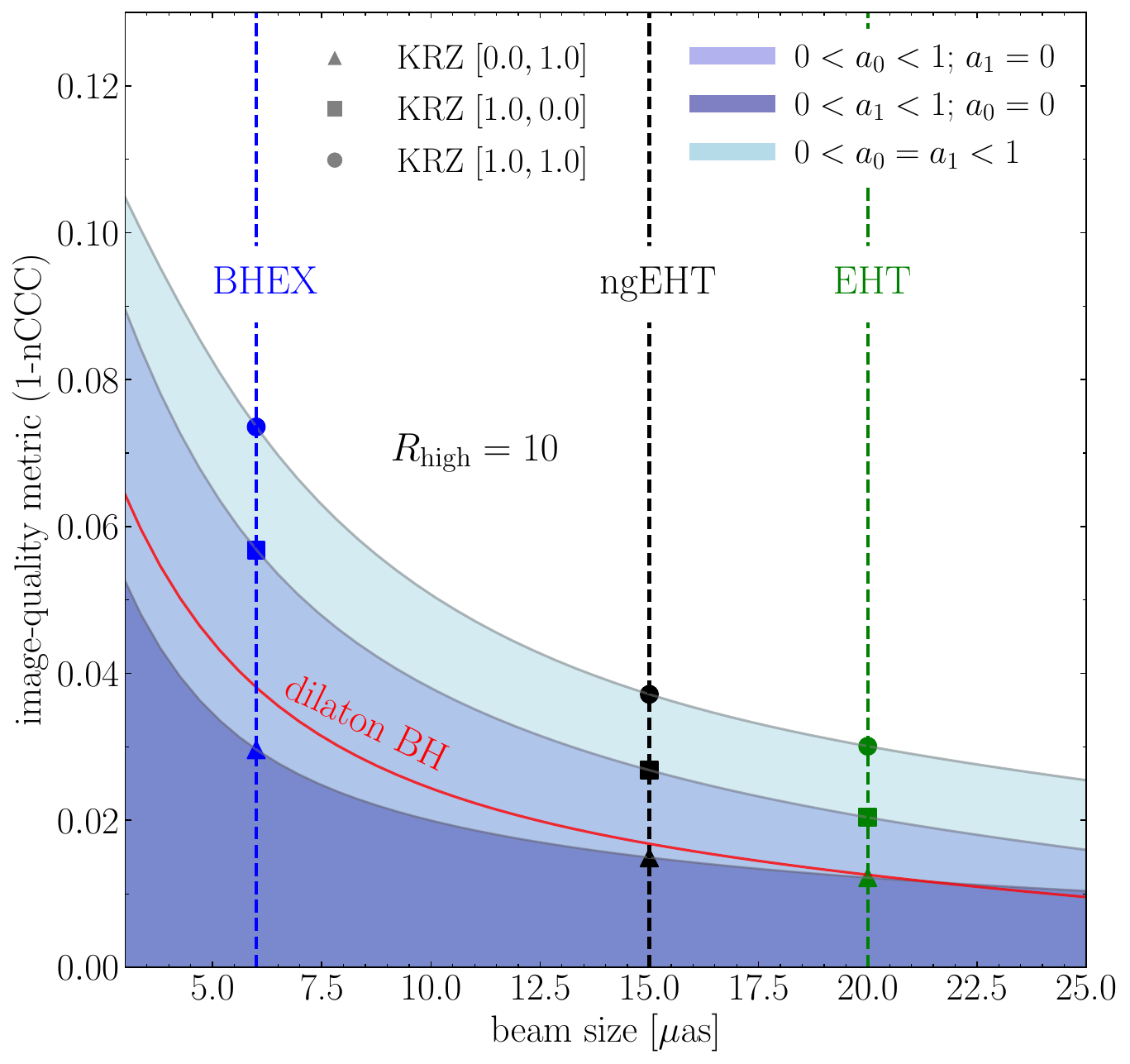}
  \caption{\textbf{Different electron-temperature parameter image-metric
      comparison.} The same as in the left panel of Fig. 5 but for
    different values of the electron-temperature parameter, namely,
    $R_{\rm high}=40$ (left panel) and $R_{\rm high}=10$ (right panel);
    by contrast, $R_{\rm high}=160$ was employed in Fig. 5.}
  \label{fig:nCCC_2}
\end{figure*}

The results of this analysis are shown in Supplementary
Fig.~\ref{grrt_2}, which is the logical equivalent of Fig. 3, but when
considering $R_{\rm high}=40$ (left panel) or $R_{\rm high}=10$ (right
panel). Similarly, Supplementary Fig.~\ref{fig:nCCC_2} can be considered
the equivalent of Fig. 5, thus evaluating the image-comparison metric in
terms of the ``mismatch'' $1-{\rm nCCC}$ for different beam sizes and KRZ
BHs and when the GRRT simulations have considered $R_{\rm high}=40$ (left
panel) or $R_{\rm high}=10$ (right panel). Both figures clearly show that
the conclusions drawn in the main text for the case $R_{\rm high}=160$,
apply both qualitatively and quantitatively irrespective of the choice
made for $R_{\rm high}$. Additional and more sophisticated approaches are
also possible, such as those involving two-temperature
plasmas~\citep{Mizuno-etal2021, Dihingia0-etal2023} or temperature
prescriptions following from first-principle particle-in-cell
calculations~\cite{Meringolo2023}. While physically more realistic and
less empirical, comparisons with the $R\!-\!\beta$ parameterisation have
highlighted that the differences are actually
small~\citep{Mizuno-etal2021, Moscibrodzka2025}, but may become important
as the precision of the observations increases.

Another degree of freedom in our analysis is represented by the
inclination angle. While we have anticipated that our results do not
depend sensitively on the inclination angle, we demonstrate this by
reporting in Supplementary Fig.~\ref{fig:grrt_60deg} information that is
the same as that in Fig. 3 (where an angle $i=30^\circ$ was employed) but
for an inclination angle of $i=60^\circ$. Clearly, much of the arguments
made in the main text for low-inclination observations apply also in the
case when the inclination is larger.

\begin{figure*}
  \centering
  \includegraphics[width=0.98\textwidth]{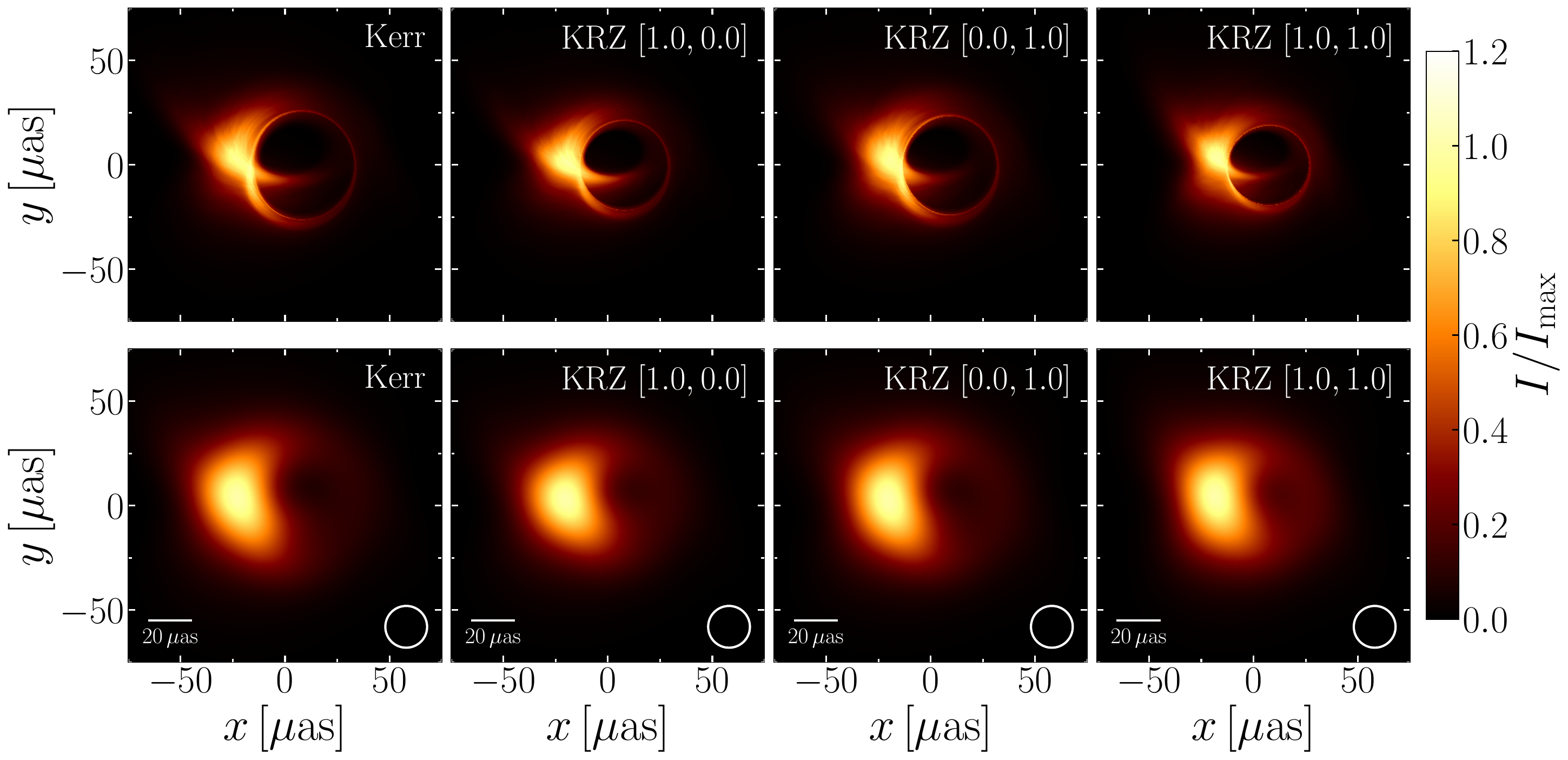}
  \caption{\textbf{Different inclination angle images.} The same as in
    Fig. 3 but for an inclination angle of $i=60^\circ$; by contrast, an
    angle $i=30^\circ$ was employed in Fig. 3.}.
    \label{fig:grrt_60deg}
\end{figure*}
\begin{figure*}
  \centering
  \includegraphics[width=0.48\textwidth]{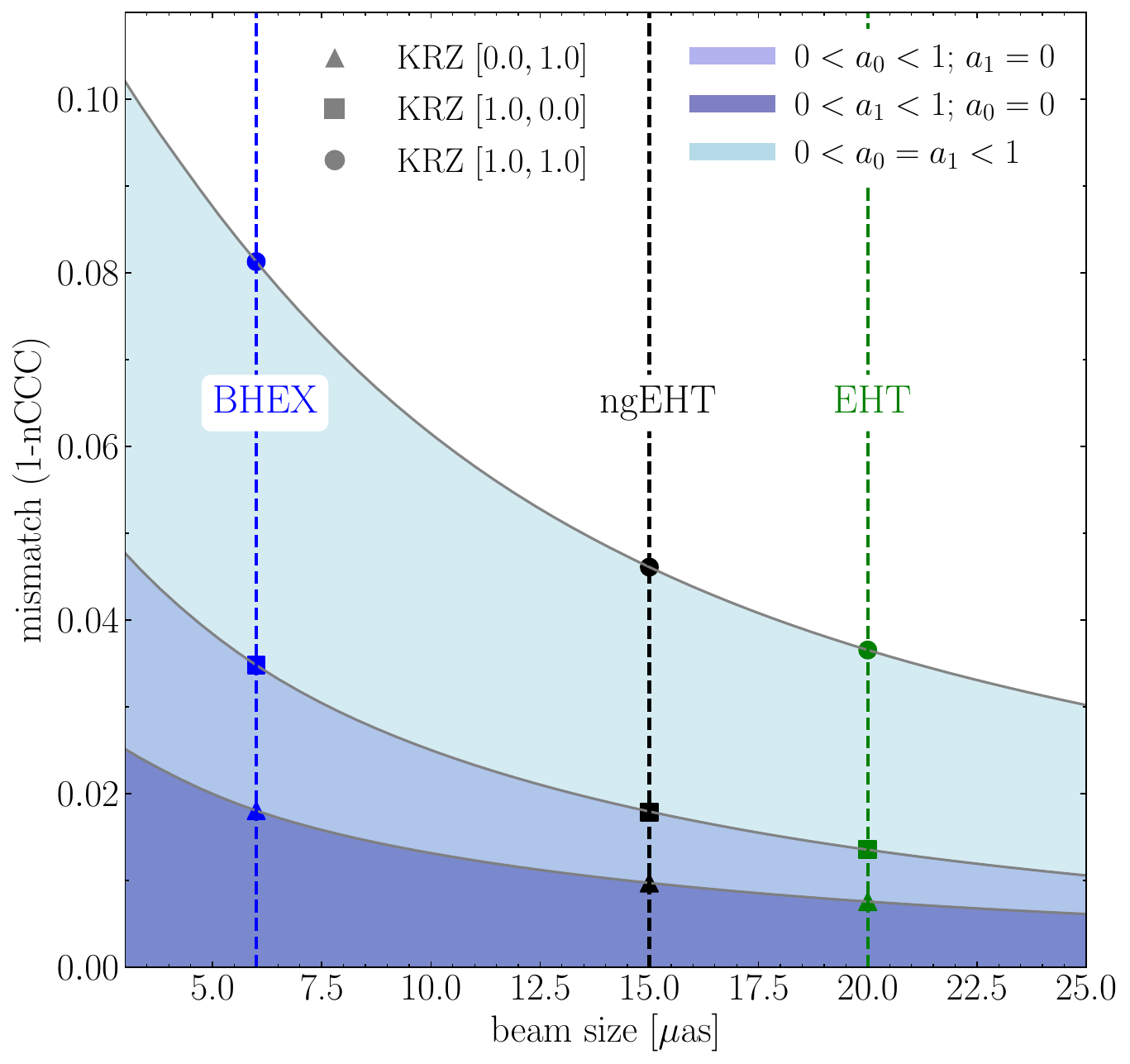}
  \hspace{0.2cm}
  \includegraphics[width=0.48\textwidth]{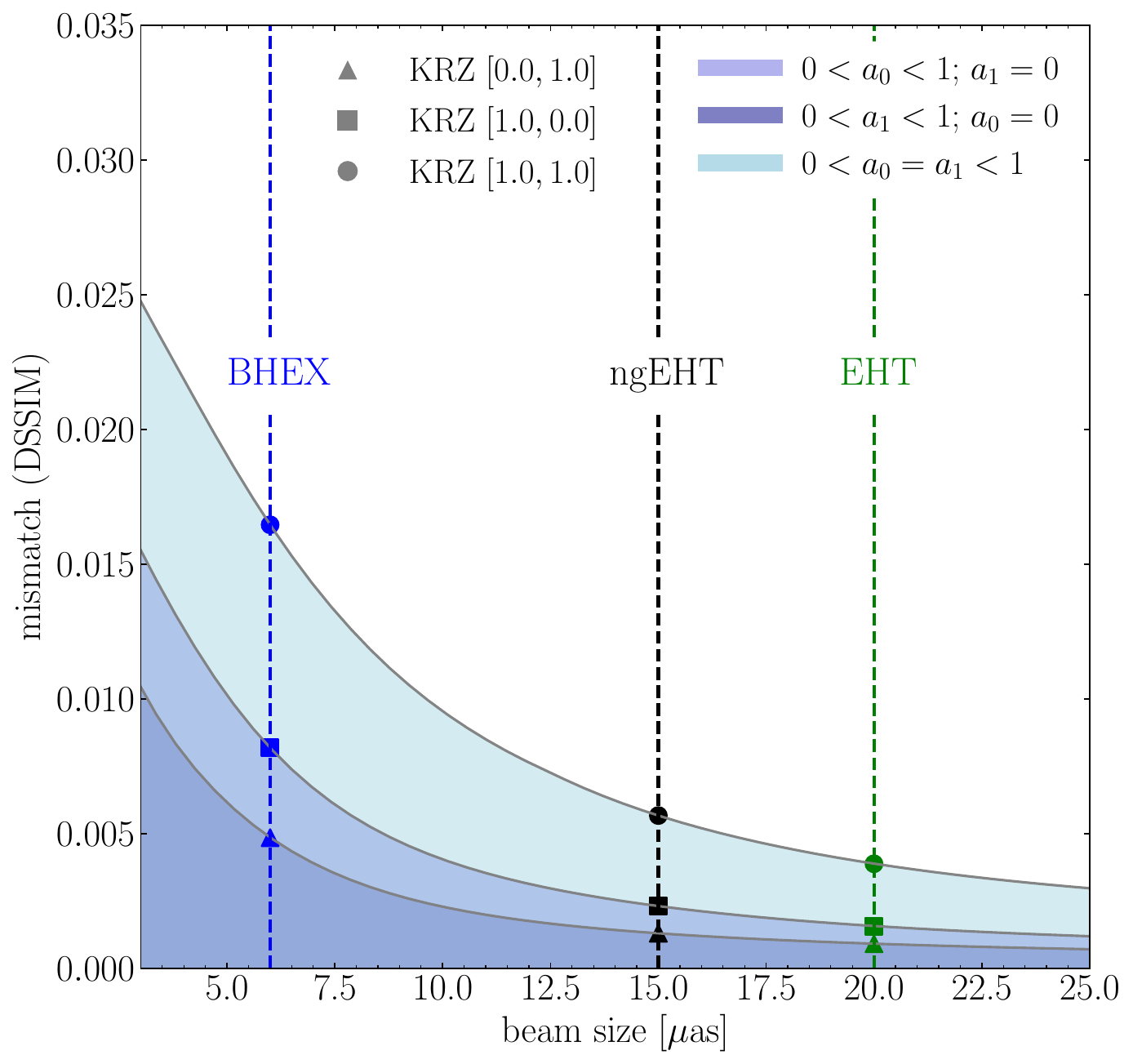}
  \caption{\textbf{Different inclination angle image-metric comparison.}
    The same as in Fig. 5 but for an inclination angle of $i=60^\circ$;
    by contrast, an angle $i=30^\circ$ was employed in Fig. 5.}
  \label{fig:nCCC_60deg}
\end{figure*}

In a similar fashion, we present in Supplementary
Fig.~\ref{fig:nCCC_60deg} the same information offered in Fig. 5 in terms
of the mismatches computed with the image comparison indices. Also in
this case, both qualitatively and quantitatively the behaviour of the
mismatches is essentially the same. Overall, the material presented in
this section provides evidence of the significant robustness of our
results.

\end{document}